\documentclass{jfm}
\usepackage{graphicx}
\usepackage{epstopdf, epsfig}

\usepackage{amsmath}   
\usepackage{txfonts}   
\usepackage{bm}        
\usepackage{subfigure} 
\usepackage{here}      
\usepackage{xr-hyper} 
\usepackage{hyperref} 
\usepackage{lineno}
\RequirePackage[normalem]{ulem} 
\RequirePackage{color}\definecolor{RED}{rgb}{1,0,0}\definecolor{BLUE}{rgb}{0,0,1} 
\providecommand{\DIFdel}[1]{{\protect\color{red}\sout{#1}}}    
\providecommand{\DIFdel}[1]{}
\usepackage{color}

\shorttitle{Spatio-temporal intermittency of the turbulent energy cascade}
\shortauthor{T. Yasuda and J. C. Vassilicos}

\title{Spatio-temporal intermittency of the turbulent energy cascade}

\author{T. Yasuda\aff{1}
  \corresp{\email{t.yasuda@imperial.ac.uk}},
  \and {J. C. Vassilicos\aff{1}
  \corresp{\email{j.c.vassilicos@imperial.ac.uk}}}}

\affiliation{\aff{1}
Turbulence, Mixing and Flow Control Group,
Department of Aeronautics, Imperial College London, UK}

\begin{document}
\maketitle



\begin{abstract}
In incompressible and periodic statistically stationary turbulence,
exchanges of turbulent energy across scales and space are
characterised by very intense and intermittent spatio-temporal
fluctuations around zero of the time-derivative term, the spatial
turbulent transport of fluctuating energy, and the pressure-velocity
term. These fluctuations are correlated with each other and with the
intense intermittent fluctuations of the interscale energy transfer
rate. These correlations are caused by the sweeping effect, the link
between non-linearity and non-locality, and also relate to geometrical
alignments between the two-point fluctuating pressure force difference
and the two-point fluctuating velocity difference in the case of the
correlation between the interscale transfer rate and the
pressure-velocity term. All these processes are absent from the
spatio-temporal average picture of the turbulence cascade in
statistically stationary and homogeneous turbulence.
\end{abstract}
\begin{keywords}
\end{keywords}

\section{Introduction}

It has been known from numerical simulations of 
turbulent flows since the early 1990s 
\citep[see][and references therein]{
piomelli_1991_subgrid-scale, 
domaradzki_1993_an,
cerutti_1998_intermittency, 
aoyama_2005_statistics,gotoa2008, 
ishihara_2009_study}
that the interscale energy transfer rate is highly intermittent.  Even
though the turbulent energy cascades, on average, from large to small
scales, the spatio-temporal fluctuations of the interscale energy
transfer rate are such that localised regions exist in the flow where
the turbulent energy actually cascades from small to large scales,
opposite to the average sense. This is often referred to as
backscatter.

In statistically stationary and homogeneous turbulence there are no
average spatial fluxes and no average time-dependence: the average
picture of turbulence interscale transfers involves only energy input
rate, interscale energy transfer rate, turbulence dissipation and
viscous diffusion in scale space. Given that the interscale transfer
rate is known to fluctuate widely about its mean, the question arises
to know what is hidden behind the average cascade picture. This
question is also raised by the fully generalised K\'arm\'an-Howarth
equation, the K\'arm\'an-Howarth-Monin-Hill (KHMH) equation without
averaging operations, because it contains a number of local spatial
fluxes and a local time-derivative term all of which can be expected
to fluctuate too. This equation is a scale-by-scale energy budget
local in space and time, directly derived from the incompressible
Navier-Stokes equations for the instantaneous velocity field
%
\citep[see][]{duchon_2000_inertial,hill_2002_exact}
without decomposition (e.g. Reynolds decomposition), without
averages (e.g. Reynolds averages), and without any assumption made
about the turbulent flow (e.g. homogeneity, isotropy, etc.).


In the next section we present this KHMH equation and the various
physical processes included in it. We then introduce in section 3 the
Direct Numerical Simulations (DNSs) of forced periodic turbulence which
we use to study the various processes involved in the KHMH
equation. The KHMH analysis of our simulations confirms the well-known
average interscale transfer picture mentioned above and the well-known
intense intermittency of the interscale energy transfer rate and
related backscatter. Having validated and set the stage for our
approach in section 3, we then proceed in section 4 with new results
concerning the fluctuations of the various processes in the KHMH
equation and their correlations. In section 5 we summarise our
conclusions and give some pointers for future research.

\section{The KHMH equation: local scale-by-scale energy budget} 
%
%
Various forms of the KHMH equation have already been used to analyse
DNS, Particle Image Velocimetry and Hot Wire Anemometry data of
various turbulent flows over the past 15 years or so 
\citep[see  e.g.][]{
marati_2004_energy,
danaila_2012_yaglom-like,
gomes-fernandes_2015_the,
togni_2015_physical, 
valente_2015_the,
cimarelli_2016_cascades,
alvesportela_2017_the}. 
The KHMH equation in its most general form, 
i.e. without averages and without assumptions about the flow, 
is the evolution equation for $|\delta \bm{u}|^2$, 
where $\delta \bm{u} \equiv \bm{u} - \bm{u}^{\prime}$ 
is the difference of fluid velocities at two separate
points $\bm{x}$ and $\bm{x}^{\prime}$, $\bm{u} \equiv
\bm{u}(\bm{x},t)$, and $\bm{u}^{\prime} \equiv
\bm{u}(\bm{x}^{\prime},t)$. It is expressed in terms of functions of
the centroid position $\bm{X} = ( \bm{x} + \bm{x}^{\prime} ) / 2$, the
separation vector $\bm{r} = \bm{x} - \bm{x}^{\prime}$ and time $t$ as
follows:
%

%
\begin{eqnarray}
\label{eq:KHMH}
\frac{\partial}{\partial t}|\delta \bm{u}|^2 &+&
\frac{\partial}{\partial r_k} (\delta u_k |\delta \bm{u}|^2 )= -
\frac{\partial}{\partial X_k} \frac{( u_k + u_k^{\prime} ) |\delta
  \bm{u}|^2}{2} - \frac{2}{\rho} \frac{\partial}{\partial X_k} (\delta
u_k \delta p ) \notag \\ &+& 2 \nu \frac{\partial^2}{\partial r_k^2}
|\delta \bm{u}|^2 + \frac{\nu}{2} \frac{\partial^2}{\partial X_k^2}
|\delta \bm{u}|^2 \notag \\ &-& \bigg[2\nu\bigg(\frac{\partial
    u_j}{\partial x_k}\bigg)^2 + 2\nu\bigg(\frac{\partial
    u^{\prime}_j}{\partial x^{\prime}_k}\bigg)^2\bigg] + 2 \delta u_k
\delta f_k
\end{eqnarray}
%
%
where $\nu$ is the kinematic viscosity, $\rho$ is the fluid density
and $\delta p = p - p^{\prime}$ and $\delta f_k = f_k - f_k^{\prime}$
are, respectively, the pressure and body force differences across the
two points ${\bm x} = {\bm X} + {\bm r}/2$ and ${\bm x}^{\prime} =
{\bm X} - {\bm r}/2$.
%
This scale-by-scale balance consists of the following eight terms:
\begin{description}
\item[(i)] $4\mathcal{A}_t(\bm{X},\bm{r},t) \equiv
  \frac{\partial}{\partial t}|\delta \bm{u}|^2$ is the time derivative
  term;
\item[(ii)] $4\Pi(\bm{X},\bm{r},t) \equiv \frac{\partial}{\partial
  r_k} (\delta u_k |\delta \bm{u}|^2)$ is the interscale energy
  transfer term;
\item[(iii)] $4\mathcal{T}(\bm{X},\bm{r},t) \equiv -
  \frac{\partial}{\partial X_k} \frac{( u_k + u_k^{\prime} ) |\delta
    \bm{u}|^2}{2}$ is the turbulent transport of $|\delta \bm{u}|^2$
  along ${\bm X}$;
\item[(iv)] $4\mathcal{T}_p(\bm{X},\bm{r},t) \equiv - \frac{2}{\rho}
  \frac{\partial}{\partial X_k} (\delta u_k \delta p )$ is the
  pressure-velocity term;
\item[(v)] $4\mathcal{D}_{\nu}(\bm{X},\bm{r},t) \equiv 
  2 \nu
  \frac{\partial^2}{\partial r_k^2} |\delta \bm{u}|^2$ is the viscous
  diffusion in the space of separation vectors $\bm{r}$;
\item[(vi)]
$4\mathcal{D}_{X,\nu}(\bm{X},\bm{r},t) \equiv 
\frac{\nu}{2} \frac{\partial^2}{\partial X_k^2} |\delta \bm{u}|^2$
is the  viscous diffusion in physical space;
\item[(vii)] $4\epsilon^{*}(\bm{X},\bm{r},t) \equiv
  \bigg[2\nu\bigg(\frac{\partial u_j}{\partial x_k}\bigg)^2 +
    2\nu\bigg(\frac{\partial u^{\prime}_j}{\partial
      x^{\prime}_k}\bigg)^2\bigg]$ is the sum of turbulent kinetic
  energy dissipations at $\bm{x}$ and $\bm{x}^{\prime}$;
\item[(viii)] $4\mathcal{I}(\bm{X},\bm{r},t) \equiv 2 \delta u_k
  \delta f_k$ is the energy input rate at separation ${\bm r}$.
\end{description}

Note that $4\Pi$, $4\mathcal{T}_p$, $4\mathcal{D}_{\nu}$,
$4\mathcal{D}_{X,\nu}$, $4\epsilon^{*}$, $4\mathcal{I}$ and
$4\mathcal{A}_t - 4\mathcal{T}$ are all Galilean invariant. The KHMH
equation (\ref{eq:KHMH}) is for fluid velocities and fluid velocity
differences, but in the particular case of homogeneous turbulence it
remains exactly the same for turbulent fluctuating velocities (which
are Galilean invariant) and turbulent fluctuating velocity differences
if written in the frame moving uniformly with the mean flow
velocity. This is a natural frame in homogeneous turbulence and
$4\mathcal{A}_t$ and $4\mathcal{T}$ are separately meaningful in this
case, as well as $4\mathcal{A}_t - 4\mathcal{T}$.


In this paper we study periodic turbulence which is forced so as to be
statistically stationary in time. We use periodic boundary conditions
because, by virtue of the Gauss divergence theorem, they set spatial
averages (over ${\bm X}$) of divergence terms such as
$4\mathcal{T}(\bm{X},\bm{r},t)$, $4\mathcal{T}_p(\bm{X},\bm{r},t)$ and
$4\mathcal{D}_{X,\nu}(\bm{X},\bm{r},t)$ to zero. Statistical
stationarity implies that the time-average of
$4\mathcal{A}_t(\bm{X},\bm{r},t)$ is also zero. Under these
conditions, the spatio-temporal average of (\ref{eq:KHMH}) reduces to
\begin{eqnarray}
\label{eq:KHMH_ST}
\langle \Pi \rangle = 
\langle \mathcal{D}_{\nu} \rangle
-\langle \epsilon^* \rangle
+\langle \mathcal{I} \rangle
\:
\end{eqnarray}
where the brackets $\langle \cdot \rangle$ signify a spatio-temporal
average (i.e. over ${\bm X}$ and $t$). As mentioned in the
introduction, the spatio-temporal average picture of turbulence
interscale transfers in statistically stationary periodic turbulence
involves only energy input rate $\mathcal{I}$, interscale energy
transfer rate $\Pi$, turbulence dissipation $\epsilon^*$ and viscous
diffusion in scale space $\mathcal{D}_{\nu}$. However, the question
naturally arises whether and to what extent the zero averages of
$4\mathcal{T}(\bm{X},\bm{r},t)$, $4\mathcal{T}_p(\bm{X},\bm{r},t)$,
$4\mathcal{D}_{X,\nu}(\bm{X},\bm{r},t)$ and
$4\mathcal{A}_t(\bm{X},\bm{r},t)$ are in any way representative of the
actual interscale and interspace energy transfer physics. These terms
are all present in the local and instantaneous balance equation
(\ref{eq:KHMH}), and if $\Pi$ fluctuates significantly, as is well
known from various previous studies, then all the other terms in this
equation must also be expected to fluctuate in both space and
time. Could it be that the spatio-temporal average, when applied to
interscale transfer and cascade dynamics, actually conceals the really
significant physics of these dynamics?

In the following section we describe our DNS of statistically
stationary periodic turbulence and validate it by confirming our
existing knowledge of the average and fluctuating interscale
transfer. In section 4 we use our DNS and the local and instantaneous
KHMH equation (\ref{eq:KHMH}) to access the dynamics which are hiding
behind the spatio-temporal averages.

\section{DNS of body-forced periodic turbulence}
%
We perform direct numerical simulations (DNSs) 
of body-forced periodic turbulence using 
a pseudo-spectral method. 
We numerically solve 
the incompressible vorticity equation
\begin{eqnarray}
\label{}
\frac{\partial \bm{\omega}}{\partial t} = 
\nabla \times (\bm{u} \times \bm{\omega}) + 
\nu \nabla^2 \bm{\omega} + \nabla \times \bm{f}
\:.
\end{eqnarray}
with the continuity equation 
\begin{eqnarray}
\label{}
\nabla \cdot \bm{u} = 0
\:,
\end{eqnarray}
where $\bm{u}$ is the velocity field, $\bm{\omega}$ is the vorticity
field, and $\bm{f}$ is the body force.
%
The boundary conditions are triply periodic, i.e. 
\begin{eqnarray}
\label{eq:bc}
\bm{u}(x_1,x_2,x_3) = \bm{u}(x_1+2\pi,x_2,x_3) =
\bm{u}(x_1,x_2+2\pi,x_3) = \bm{u}(x_1,x_2,x_3+2\pi)
\:.
\end{eqnarray}

\begin{figure}
\centering
\subfigure{
        \begin{minipage}{0.5\linewidth}
                \includegraphics[clip,width=\linewidth]
                {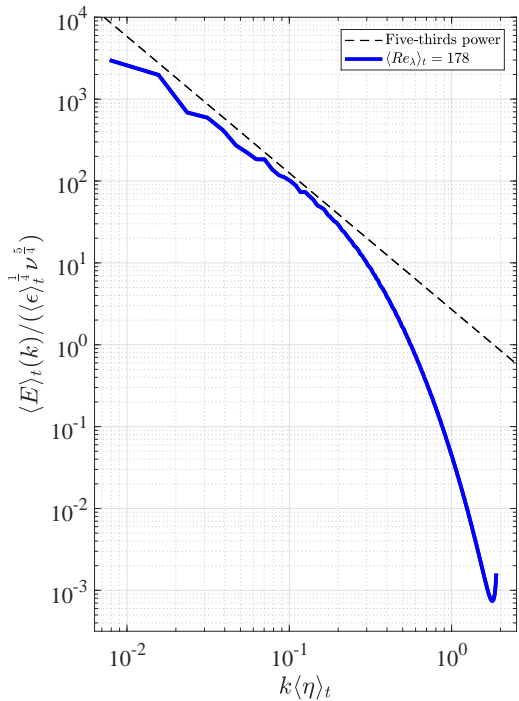}
        \end{minipage}
}
\caption{Time-average 3D energy spectrum of forced periodic
  turbulence. The dashed line indicates the $-5/3$ slope for
  reference.}
\label{fig:spectra}
\end{figure}

The forcing method employed here is a negative damping forcing
\citep[see
  e.g.][]{mccomb_2015_self-organization,linkmann_2015_sudden}:
\begin{eqnarray}
\label{eq:mccombforce}
\tilde{\bm{f}}(\bm{k},t) 
&=& (\epsilon_W/2K_f) \tilde{\bm{u}}(\bm{k},t), 
\:\:\:\:\:\: {\rm for} 
\:\:\: 0 < |\bm{k}| < k_f, \notag \\
&=& 0, 
\:\:\:\:\:\:\:\:\:
\:\:\:\:\:\:\:\:\:\:\:\:\:\:\:\:\:
\:\:\:\:\:\:\:\:\:\:\:\:\:\:\:\:\:  {\rm otherwise}
\:,
\end{eqnarray}
where $\tilde{\bm{f}}(\bm{k},t)$ and $\tilde{\bm{u}}(\bm{k},t)$ are
the Fourier coefficients of ${\bm{f}}(\bm{x},t)$ and
${\bm{u}}(\bm{x},t)$ respectively, $\epsilon_W$ is the energy input
rate per unit mass, $k_f$ is the cutoff wavenumber, and $K_f$ is the
total kinetic energy per unit mass contained in the forcing band $0 <
|\bm{k}| < k_f$. (We have also tried various other forcing methods
without significant changes to our results all of which involve an
average over two-point orientations as explained in the penultimate
paragraph of this section and are therefore insensitive to any
potential anisotropies introduced by forcing.)
In our simulations, we set $\epsilon_W=0.1$ and $k_f=2.5$.
While $\epsilon_W$ is set constant in time, the total kinetic energy
per unit mass $K(t)$ and the total energy dissipation per unit mass
$\epsilon(t)$ fluctuate around constant values.
We run two simulations and therefore use two mesh sizes, $128^3$ and
$512^3$, with respective kinematic viscosities $\nu=0.003$ and
$0.00072$.
Using these parameters, the simulations are run for about 82 and 35
eddy turnover times $\langle L \rangle_{t}/\sqrt{{2\over 3} \langle K
  \rangle_{t}}$ respectively, where $L$ is the integral length-scale
calculated from the 3D energy spectrum $E(k,t)$, i.e. $L(t) =
{3\pi\over 4} \int_{0}^{\infty} k^{-1} E(k,t) dk/K(t)$. Throughout
this work, suitably converged time-averages $\langle . \rangle_{t}$
are calculated over 30 eddy turnover times for the $128^3$ simulation
and over 8.63 eddy turnover times for the $512^3$ simulation,
i.e. from $t = 25 \langle L \rangle_{t}/\sqrt{{2\over 3} \langle K
  \rangle_{t}}$ to $t = 55 \langle L \rangle_{t}/\sqrt{{2\over 3}
  \langle K \rangle_{t}}$ for $128^3$ and from $t = 8.86 \langle L
\rangle_{t}/\sqrt{{2\over 3} \langle K \rangle_{t}}$ to $t = 17.49
\langle L \rangle_{t}/\sqrt{{2\over 3} \langle K \rangle_{t}}$ for
$512^3$.  The time-averaged Taylor length-scale Reynolds number (based
on the Taylor length $\lambda = \sqrt{10\nu K/\epsilon}$) is $\langle
Re_{\lambda} \rangle_t = 80.9$ and $\langle Re_{\lambda} \rangle_t =
178$ respectively. The average spatial resolution $k_{max}\langle \eta
\rangle_t$ equals $1.37$ for $128^3$ and $1.89$ for $512^3$ with
respective standard deviations $0.0189$ and $0.0426$, 
where $\eta$ is the Kolmogorov length. 
The time-average integral length-scale is related to $k_f$ by $2\pi / k_f = 2.28
\langle L \rangle_t$ and $2\pi / k_f = 2.35 \langle L \rangle_t$
respectively and the standard deviations of $\lambda$ and $L$ are
3.4\% of $\langle \lambda \rangle_{t}$ and 5.9\% of $\langle L
\rangle_{t}$ in the $128^3$ case and 5.2\% of $\langle \lambda
\rangle_{t}$ and 7.8\% of $\langle L \rangle_{t}$ in the $512^3$ case.
%
In Fig.~\ref{fig:spectra} the 3D energy spectrum is plotted for the
$512^3$ case.
%

\begin{figure}
\centering
\subfigure{
        \begin{minipage}{0.7\linewidth}
                \includegraphics[clip,width=\linewidth]
                {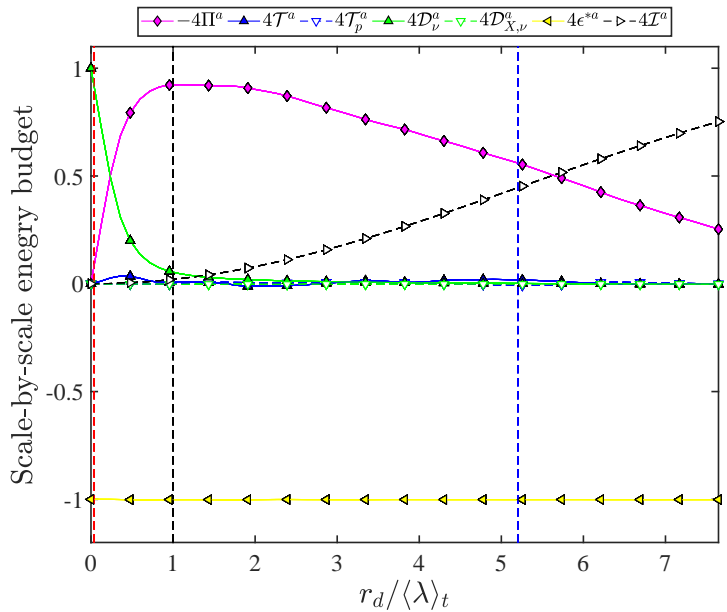}
        \end{minipage}
}
\caption{Average scale-by-scale energy budget for statistically
  stationary periodic turbulence ($\langle Re_{\lambda} \rangle_t =
  178$): plots of $\langle Q^{a} \rangle (r_d )$ normalised by
  $\langle 4\epsilon^{*a} \rangle$ versus $r_{d}/\langle \lambda
  \rangle_{t}$ for the terms, generically referred to as $Q$ in this
  caption, in (\ref{eq:KHMH}). The actual list of these terms is at
  the top of the plot. Vertical red, black, blue dashed lines
  denote the Kolmogorov scale $\langle \eta \rangle_t$, the
  Taylor-micro scale $\langle \lambda \rangle_t$, and the integral
  scale $\langle L \rangle_t$, respectively. Similar results (not
  plotted here) are obtained for $\langle Re_{\lambda} \rangle_t =
  80.9$ except that the maximum value of $-4\Pi^a$ is smaller and the
  decrease of $-4\Pi^a$ from $r_{d}=\langle \lambda \rangle_t$ to
  $r_{d}=\langle L \rangle_t$ is steeper because this range of scales
  is also smaller ($\langle L \rangle_{t} / \langle \lambda
  \rangle_{t} \approx 2.7$ for $\langle Re_{\lambda} \rangle_t =
  80.9$).}
\label{fig:khmh_balance}
\end{figure}

We now use our DNS to illustrate the average scale-by-scale energy
budget (\ref{eq:KHMH_ST}). As this budget holds for arbitrary
separation vector ${\bm r}$, we calculate surface averages over
spheres of diameter $r_d$ in $\bm{r}$-space
and define $Q^a (r_d ) \equiv (\pi r_d^2)^{-1} \iiint
\limits_{|\bm{r}| = r_d} Q(\bm{r}){\rm d}\bm{r}$ for any term $Q$ in
(\ref{eq:KHMH}). In figure~\ref{fig:khmh_balance} we plot all the
spatio-temporal average terms $\langle Q^{a} \rangle (r_d )$
normalised by $\langle 4\epsilon^{*a} \rangle$ as functions of $r_{d}/
\langle \lambda \rangle_{t}$, where $\langle \lambda \rangle_{t}$ is
the time-average Taylor microscale. The only non-zero terms are indeed
those present in the average balance (\ref{eq:KHMH_ST}). We recover
the kinematic constraints (i) $\langle \mathcal{D}_{\nu}^{a} \rangle =
\langle \epsilon^{*a} \rangle$ at $r_{d}=0$ and (ii) that $\langle
\mathcal{D}_{\nu}^{a} \rangle$ is negligible compared to $\langle
\epsilon^{*a} \rangle$ for $r_{d}$ smaller than the Taylor microscale
\citep{valente_2015_the}. At scales $r_d$ larger than the Taylor
microscale, $\langle \Pi^{a} \rangle \approx -\langle \epsilon^{*a}
\rangle + \langle \mathcal{I}^{a} \rangle$, see (\ref{eq:KHMH_ST}),
and one can see from figure~\ref{fig:khmh_balance} that the direct
influence of the large-scale energy input rate diminishes as $r_d$
decreases below the integral length-scale and towards the Taylor
micro-scale. One might imagine that, as the Reynolds number increases
above the values in the present DNS, the direct influence of the
large-scale forcing might disappear in much of the inertial range of
scales between $\langle \lambda \rangle_t$ and $\langle L \rangle_t$
leading to $\langle \Pi^{a} \rangle \approx -\langle \epsilon^{*a}
\rangle$ in much of that range. This approximate equality encapsulates
the Kolmogorov equilibrium cascade. This is an average cascade where
the turbulent energy cascades from large to small inertial scales at
constant rate approximately equal to the dissipation rate.


\begin{figure}
\centering
\subfigure{
       \begin{minipage}{0.7\linewidth}
                \centerline{(a)}
                \includegraphics[clip,width=\linewidth]
                {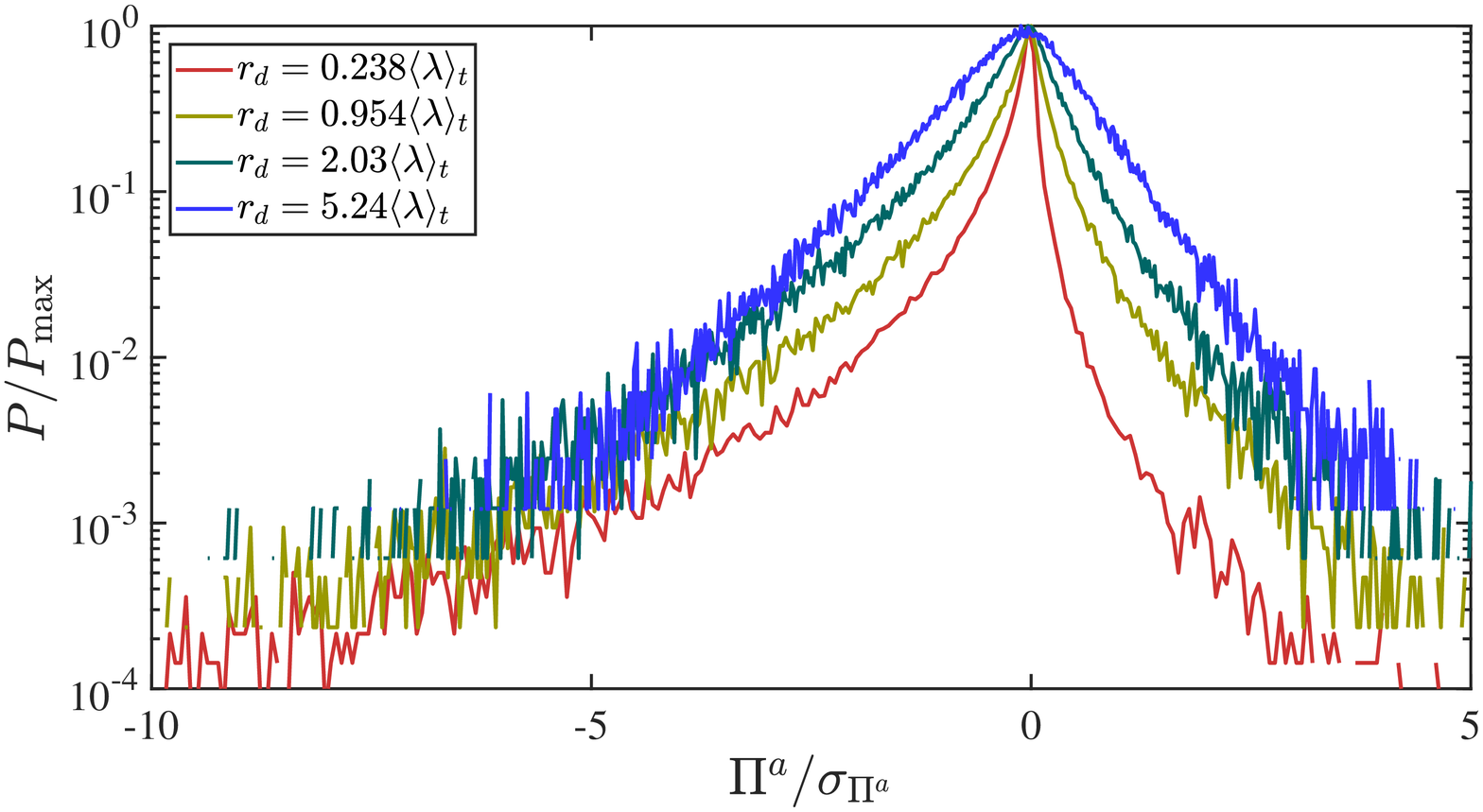}
        \end{minipage}
}
\subfigure{
        \begin{minipage}{0.7\linewidth}
                \centerline{(b)}
                \includegraphics[clip,width=\linewidth]
                {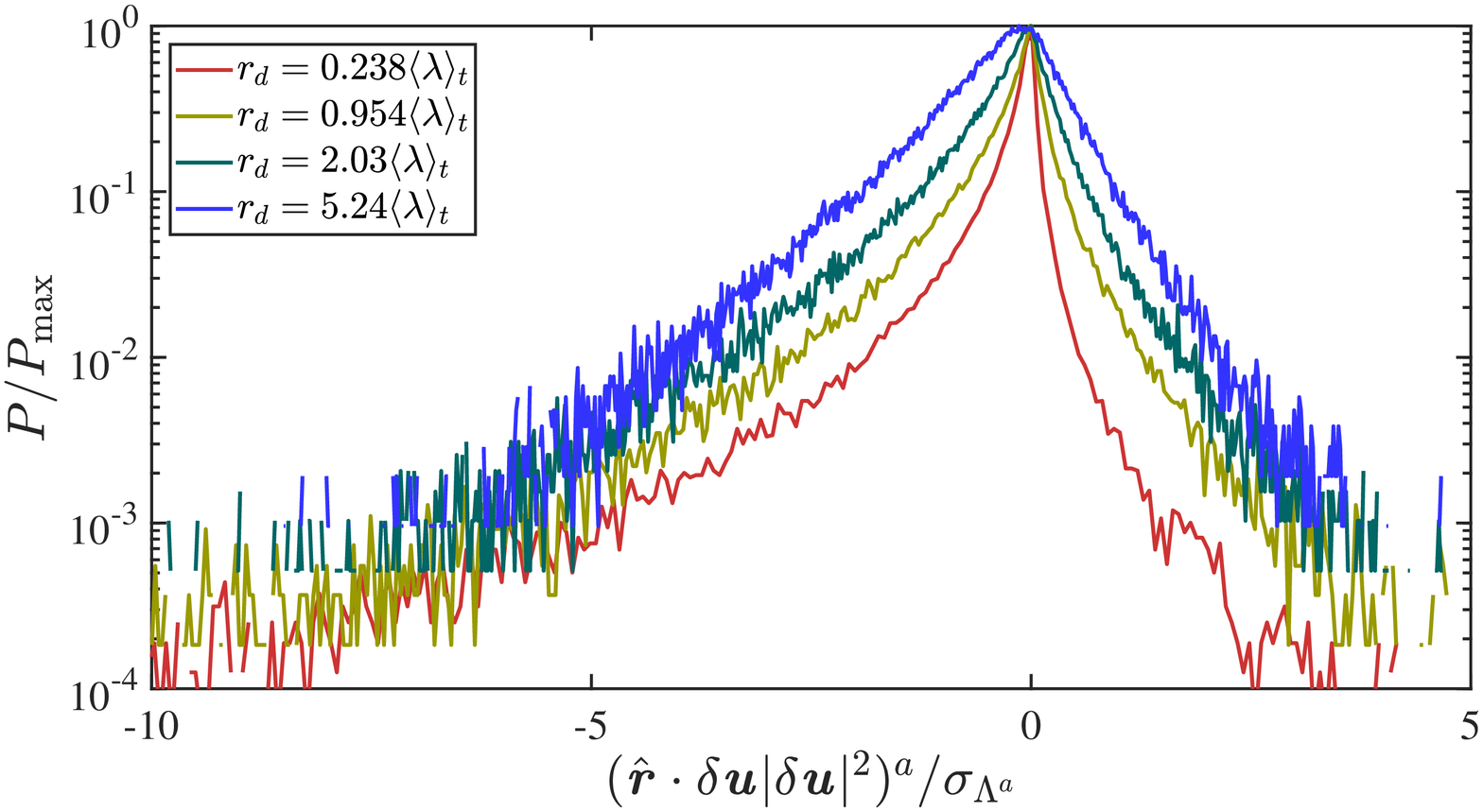}
        \end{minipage}
}
\caption{
(a) PDFs of $\Pi^a$ obtained by sampling through both space and time
  and plotted versus $\Pi^a/\sigma_{\Pi^{a}}$ where $\Pi^a (r_d )
  \equiv (\pi r_d^2)^{-1} \iiint \limits_{|\bm{r}| = r_d}
  \Pi(\bm{r}){\rm d}\bm{r}$ and $\sigma_{\Pi^{a}}$ is the standard
  deviation of $\Pi^a$. (b) PDFs of $(\bm{\hat{r}}\cdot \delta \bm{u}
  \vert \delta \bm{u} \vert^{2})^{a} \equiv (\pi r_d^2)^{-1} \iiint
  \limits_{|\bm{r}| = r_d} \bm{\hat{r}}\cdot \delta \bm{u} \vert
  \delta \bm{u} \vert^{2} {\rm d}\bm{r}$ obtained by sampling through
  both space and time and plotted versus 
  $(\bm{\hat{r}}\cdot \delta \bm{u} \vert \delta \bm{u} 
  \vert^{2})^{a}/\sigma_{\Lambda^{a}}$ where $\sigma_{\Lambda^{a}}$ 
  is the standard deviation of $(\bm{\hat{r}}\cdot \delta \bm{u} 
  \vert \delta \bm{u} \vert^{2})^{a}$. $\langle Re_{\lambda} 
  \rangle_t = 178$ but the
  plots are similar for $\langle Re_{\lambda} \rangle_t = 80.9$.
Red, yellow, green and blue lines correspond to
$r_d = 0.238 \langle \lambda \rangle_t$,
$0.954 \langle \lambda \rangle_t$,
$2.03 \langle \lambda \rangle_t$ and
$5.24 \langle \lambda \rangle_t$,
where $\lambda$ is the Taylor-micro scale and
$\langle \cdot \rangle_t$ denotes a long-term average.
%
%
}
\label{fig:pdf_TT}
\end{figure}

This average picture is punctuated by the known fact 
\citep[][]{ piomelli_1991_subgrid-scale, domaradzki_1993_an,
  cerutti_1998_intermittency, aoyama_2005_statistics,gotoa2008,
  ishihara_2009_study} that the interscale energy transfer rate $\Pi$
is highly intermittent, as indeed confirmed by our DNS. In
figure~\ref{fig:pdf_TT} we plot the PDFs of $\Pi^a$ and of
$(\bm{\hat{r}}\cdot \delta \bm{u} \vert \delta \bm{u} \vert^{2})^{a}$
for two length-scales $r_{d}$ in the approximately inertial range of
our simulation, $r_{d}/\langle \lambda \rangle_{t} \approx 2$ and $5$,
and two length-scales $r_{d}$ in the dissipation range at and below
$\langle \lambda \rangle_{t}$. Figure 3(a) is qualitatively similar to
figure 6 in \citet{ishihara_2009_study} and, along with figure 3(b),
reveals very clearly the highly intermittent and non-Gaussian nature
of $\Pi^{a}$ and $(\bm{\hat{r}}\cdot \delta \bm{u} \vert \delta \bm{u}
\vert^{2})^{a}$ given the very heavy tails of their PDFs. The equality
between the volume integral $\iiint\limits_{|\bm{r}| \le r_d}
4\Pi(\bm{X},\bm{r},t) d{\bm r}$ and the surface integral
$\iiint\limits_{|\bm{r}| = r_d} {\delta {\bm u}}\cdot \hat{\bm r}
|{\delta {\bm u}}|^{2} d{\bm r}$ implies that compression events where
$\delta {\bm u}\cdot \hat{\bm r} <0$ contribute to the forward cascade
from large to small scales whereas stretching events where $\delta
{\bm u}\cdot \hat{\bm r} >0$ contribute to backscatter, i.e. from
small to large scales. We also found a significant positive
correlation between $\Pi^a$ and $(\bm{\hat{r}}\cdot \delta \bm{u}
\vert \delta \bm{u} \vert^{2})^{a}$; their correlation coefficient
(calculated by averaging over both space and time) is close to 1 near
$r_{d}=0$ and decreases gently with increasing $r_d$ reaching values
approximately equal to 0.9 at $r_d \approx \langle \lambda \rangle_t$
and 0.6 at $r_d \approx \langle L \rangle_t$ (for $\langle
Re_{\lambda} \rangle_t = 178$). Hence, positive/negative values of
$\Pi^a$ correlate with forward/backscatter cascade events and figure 3
confirms that backscatter and extreme forward cascade events
exist with much higher probability than for a normal distribution. In
fact, these rare but powerful forward and inverse cascade events are
so very significant that 50\% of the average values
$\langle\Pi^a\rangle$ and $\langle(\bm{\hat{r}}\cdot \delta \bm{u}
\vert \delta \bm{u} \vert^{2})^{a}\rangle$ are contributed by events
whose probabilities are lower than $0.0157 P^{\Pi}_{max}$ and
$0.0145P^{\Lambda}_{max}$ respectively at $r_{d} =
0.977\langle\lambda\rangle_{t}$;
lower than $0.0350 P^{\Pi}_{max}$ and $0.0299 P^{\Lambda}_{max}$
respectively at $r_{d} = 2.03\langle\lambda\rangle_{t}$; and lower
than $0.0801 P^{\Pi}_{max}$ and $0.0738 P^{\Lambda}_{max}$
respectively at $r_{d} = 5.24\langle\lambda\rangle_{t}$ which is
close to $\langle L \rangle_{t}$. These probabilities are given for
$\langle Re_{\lambda} \rangle_t = 178$ and are even lower than for
$\langle Re_{\lambda} \rangle_t = 80.9$ (e.g. for $\langle
Re_{\lambda} \rangle_t = 80.9$, 50\% of the average value
$\langle\Pi^a\rangle$ is contributed by events whose probabilities are
lower than $0.0235 P^{\Pi}_{max}$ if $r_{d} = 0.977
\langle\lambda\rangle_{t}$ and lower than $0.0467 P^{\Pi}_{max}$ if
$r_{d} = 1.95 \langle\lambda\rangle_{t}$). $P^{\Pi}_{max}$ and
$P^{\Lambda}_{max}$ are the probabilities of the most likely events
which are, respectively, $\Pi^a =0$ and $(\bm{\hat{r}}\cdot \delta
\bm{u} \vert \delta \bm{u} \vert^{2})^{a} = 0$, i.e. no interscale
energy transfer whatsoever. The asymmetry of the PDFs in
figure~\ref{fig:pdf_TT} ensures that $\langle\Pi^a\rangle$ is negative
on average and that the average picture is therefore one of forward
interscale energy transfer. However, this average picture {\it
  requires} extreme events, both forward and backward, to fully
emerge. As mentioned in the introduction, the extreme intermittency of
$\Pi^{a}$ raises the question of the fluctuations of all the other
terms in the KHMH equation, particularly those which average out to
zero and could therefore be thought of as unimportant. Having
presented and validated our DNS and KHMH approach in this section, we
now move to the new results of this paper.

\begin{figure}
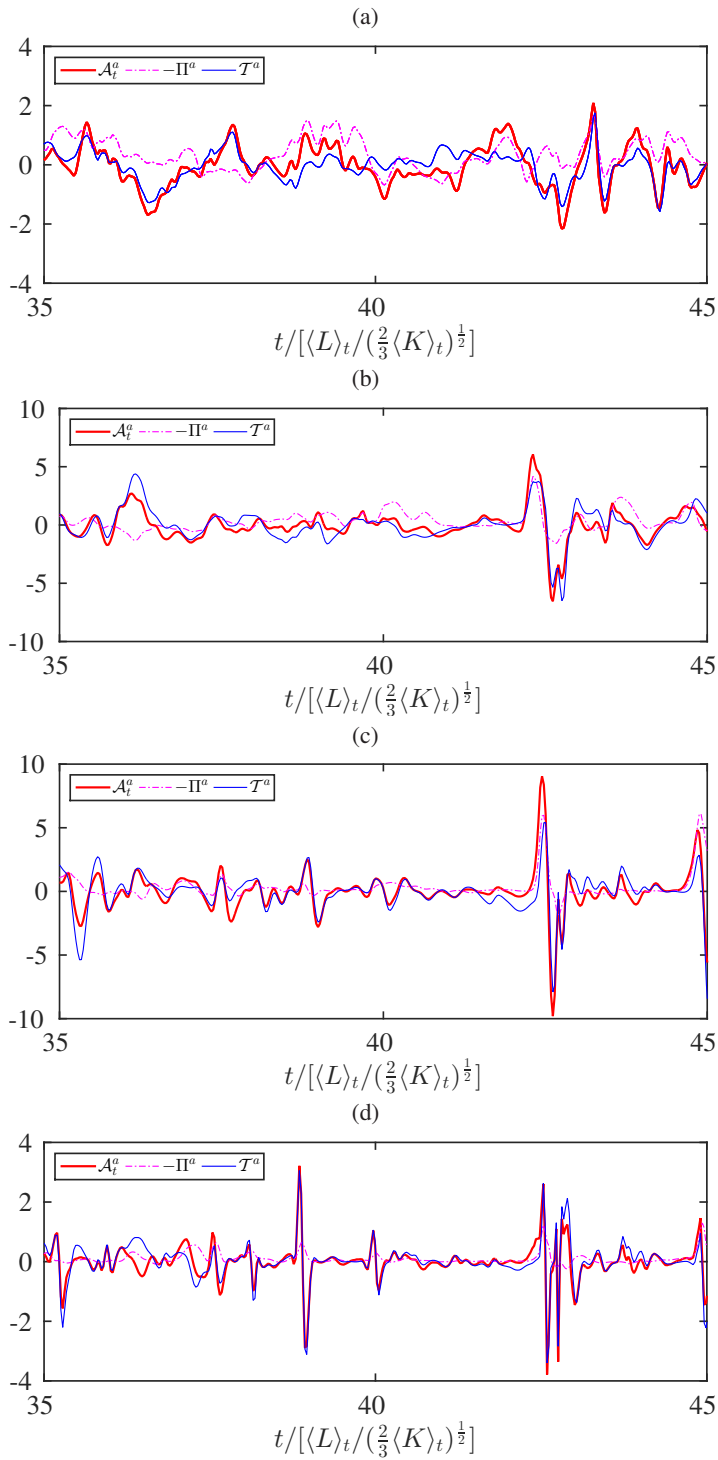

\centering
\subfigure{
       \begin{minipage}{0.7\linewidth}
                \centerline{(a)}
                \includegraphics[clip,width=\linewidth]
                {FIG4a.eps}
                \centerline{(b)}
                \includegraphics[clip,width=\linewidth]
                {FIG4b.eps}
                \centerline{(c)}
                \includegraphics[clip,width=\linewidth]
                {FIG4c.eps}
                \centerline{(d)}
                \includegraphics[clip,width=\linewidth]
                {FIG4d.eps}
       \end{minipage}
}
\caption{
Time evolutions of $\mathcal{A}_{t}^{a}$, $-\Pi^a$ and $\mathcal{T}^a$
at a random location ${\bm X}$ and four different scales. $\langle
Re_{\lambda} \rangle_t = 80.9$.
Red, magenta, blue lines represent $\mathcal{A}_{t}^a$, $-\Pi^a$,
$\mathcal{T}^a$, respectively.
(a) $r_d = 3.91\langle \lambda \rangle_t$, 
(b) $1.95 \langle \lambda \rangle_t$, 
(c) $0.977 \langle \lambda \rangle_t$, 
(d) $0.244 \langle \lambda \rangle_t$. 
}
\label{fig:timeevol_KHMH}
\end{figure}

\section{Fluctuating and correlated interscale and interspace energy transfer dynamics}
%
In figure~\ref{fig:timeevol_KHMH} we plot representative examples of
time evolutions of some orientation-averaged KHMH terms at a randomly
selected location ${\bm X}$, specifically $\mathcal{A}_{t}^{a}$,
$\mathcal{T}^a$ and $-\Pi^a$, for four different values of $r_d$ ($r_d
/ \langle \lambda \rangle_t = 3.91 , 1.95 , 0.977 , 0.224$). We chose
to plot only a few of the zero-average KHMH terms for clarity of
exposition and we plot them alongside $-\Pi^a$ to compare
fluctuations. We also chose to show plots in
figure~\ref{fig:timeevol_KHMH} which have been obtained for $\langle
Re_{\lambda} \rangle_t = 80.9$ but very similar plots are also
obtained for $\langle Re_{\lambda} \rangle_t = 178$. Plots of the same
terms at a randomly selected time $t$ but as functions of $X_1$, $X_2$
or $X_3$ rather than as functions of $t$ for a randomly selected ${\bm
  X} = (X_{1}, X_{2}, X_{3})$, look similar to those in
figure~\ref{fig:timeevol_KHMH} and lead to the same observations.

At the largest $r_d$ in figure~\ref{fig:timeevol_KHMH} 
(which is 1.47 $\langle L \rangle_{t}$) 
the fluctuations of all three quantities are comparable and one can even
detect a correlation between them
(figure~\ref{fig:timeevol_KHMH}(a)). With decreasing $r_d$, all three
signals become increasingly intermittent; all three PDFs exhibit
increasingly heavy tails (not all shown here for economy of space)
with decreasing $r_d$ as in figure~\ref{fig:pdf_TT}, except that the
PDFs of $\mathcal{A}_{t}^{a}$ and $\mathcal{T}^a$ are symmetric
whereas the PDF of $-\Pi^a$ is skewed (see figure~\ref{fig:pdf_TT}).
The fluctuations of $\mathcal{A}_{t}^{a}$ and $\mathcal{T}^a$ grow in
magnitude with decreasing $r_d$ whereas those of $-\Pi^a$ do not so
significantly. Furthermore, the correlation between
$\mathcal{A}_{t}^{a}$ and $\mathcal{T}^a$ strengthens whereas the
correlation of $-\Pi^a$ with the other two signals weakens.

\begin{figure}
\centering
\subfigure{
        \begin{minipage}{0.7\linewidth}
                \centerline{(a)}
                \includegraphics[clip,width=\linewidth]
                {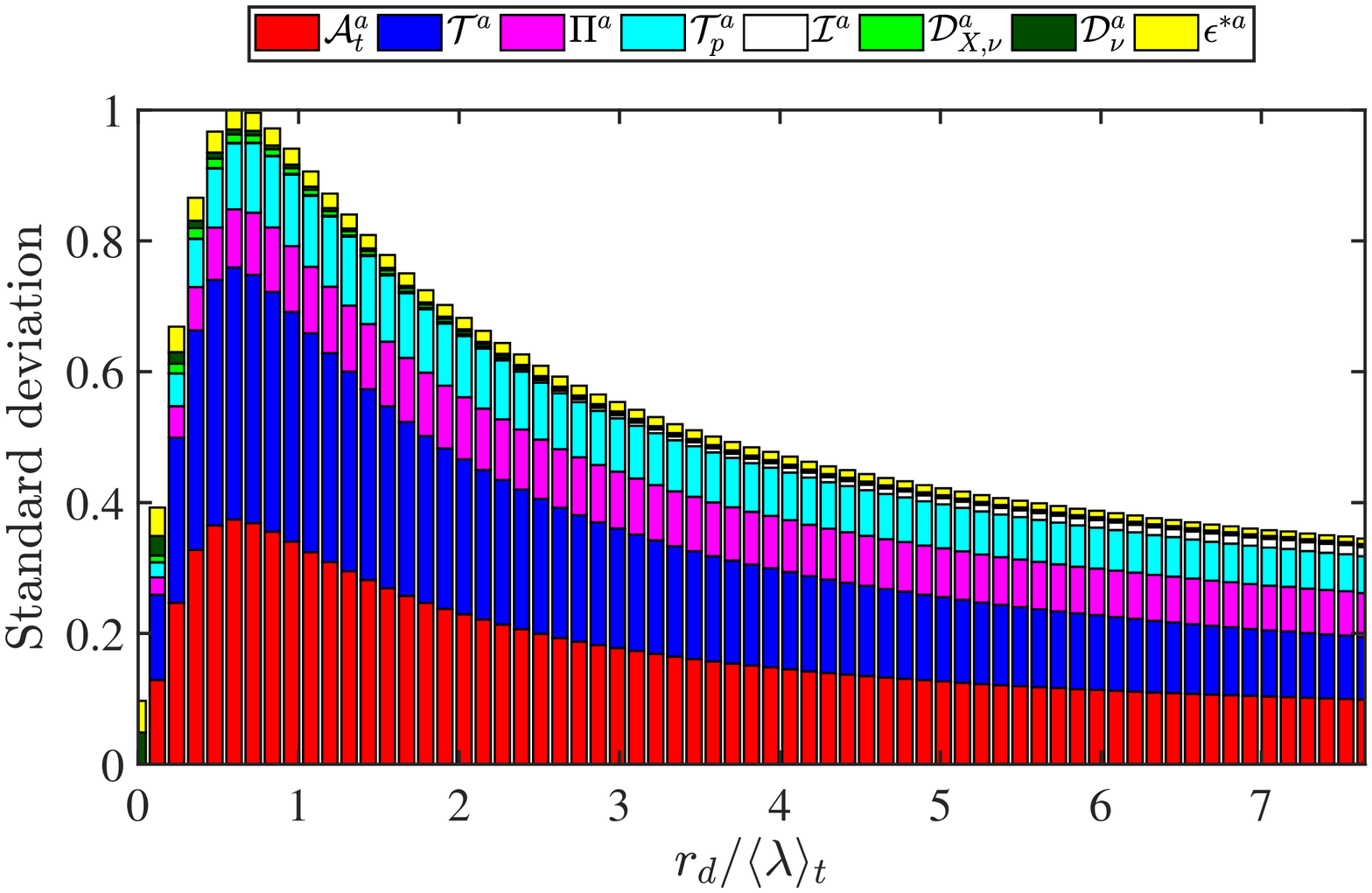}
        \end{minipage}
}
\subfigure{
        \begin{minipage}{0.7\linewidth}
                \centerline{(b)}
                \includegraphics[clip,width=\linewidth]
                {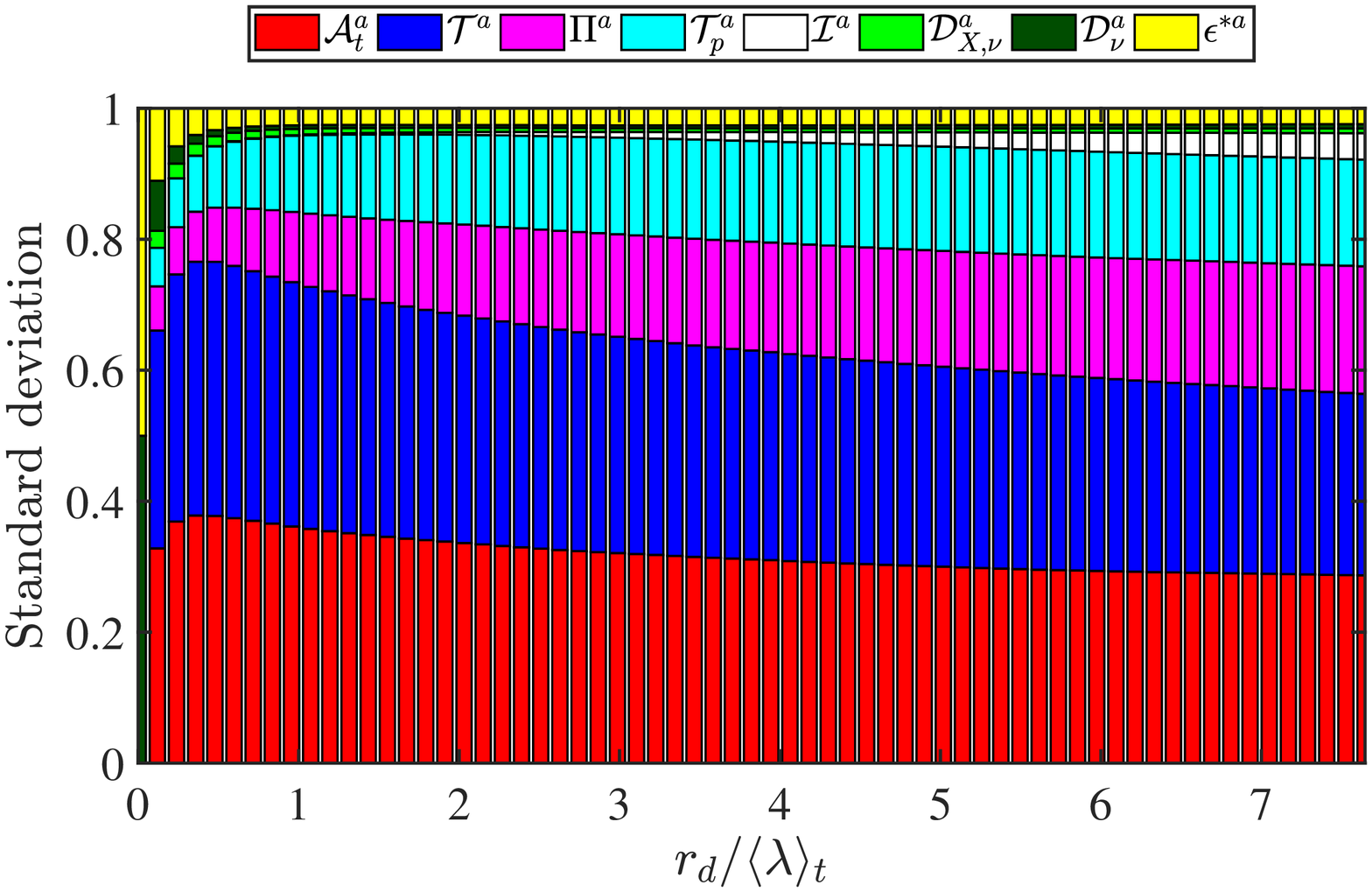}
        \end{minipage}
}
\caption{Stacked column chart of standard deviations of KHMH terms as
  a function of $r_d/\langle \lambda \rangle_t$.  Vertical axis is
  normalized by (a) the total height of the largest stacked column and
  (b) the total height of each stacked column. $\langle Re_{\lambda}
  \rangle_t = 178$ but similar results are obtained for $\langle
  Re_{\lambda} \rangle_t = 80.9$.}
\label{fig:deviations}
\end{figure}

\begin{figure}
\centering
\subfigure{
        \begin{minipage}{0.5\linewidth}
                \centerline{(a1)}
                \includegraphics[clip,width=\linewidth]
                {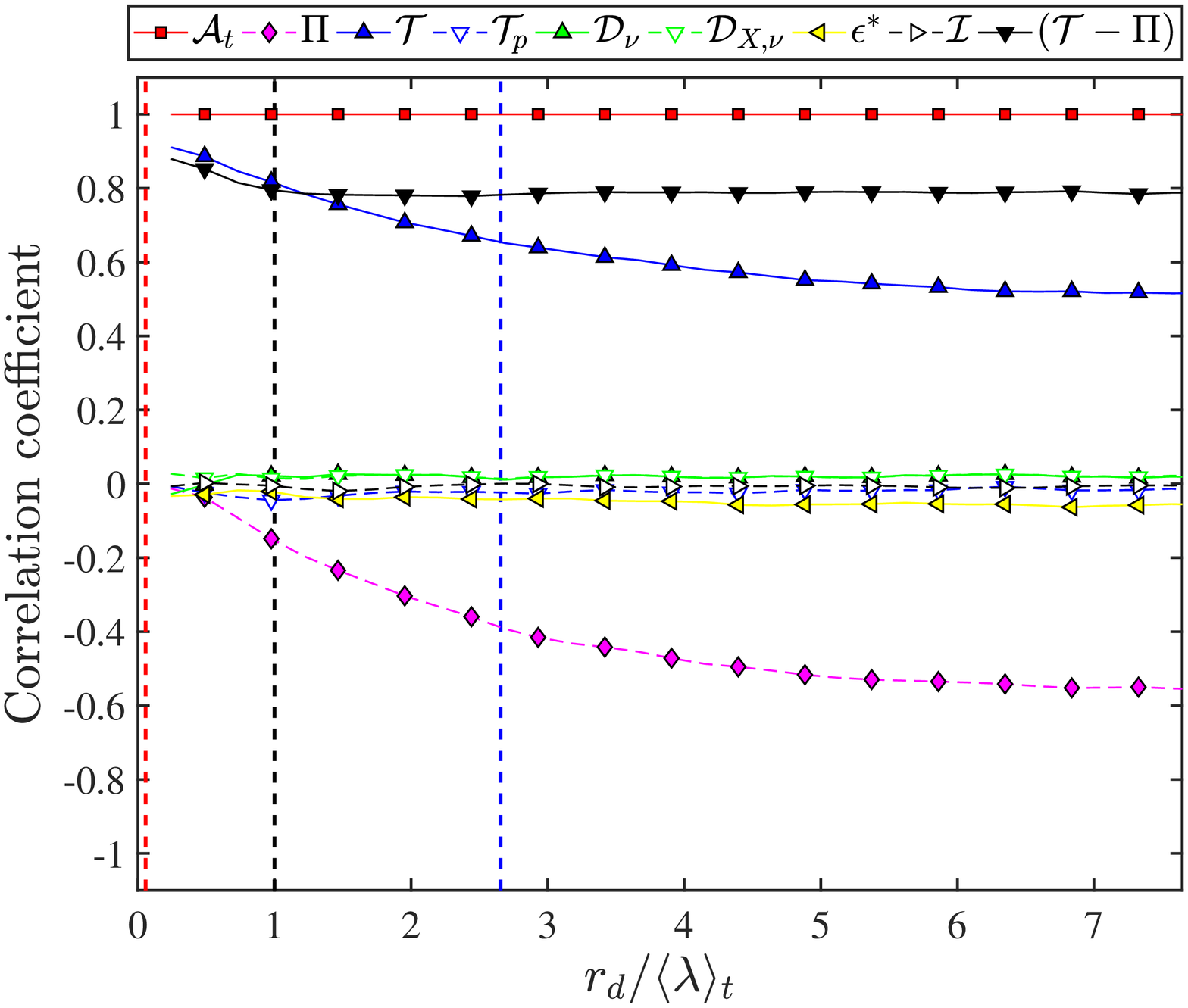}
        \end{minipage}
        \begin{minipage}{0.5\linewidth}
                \centerline{(b1)}
                \includegraphics[clip,width=\linewidth]
                {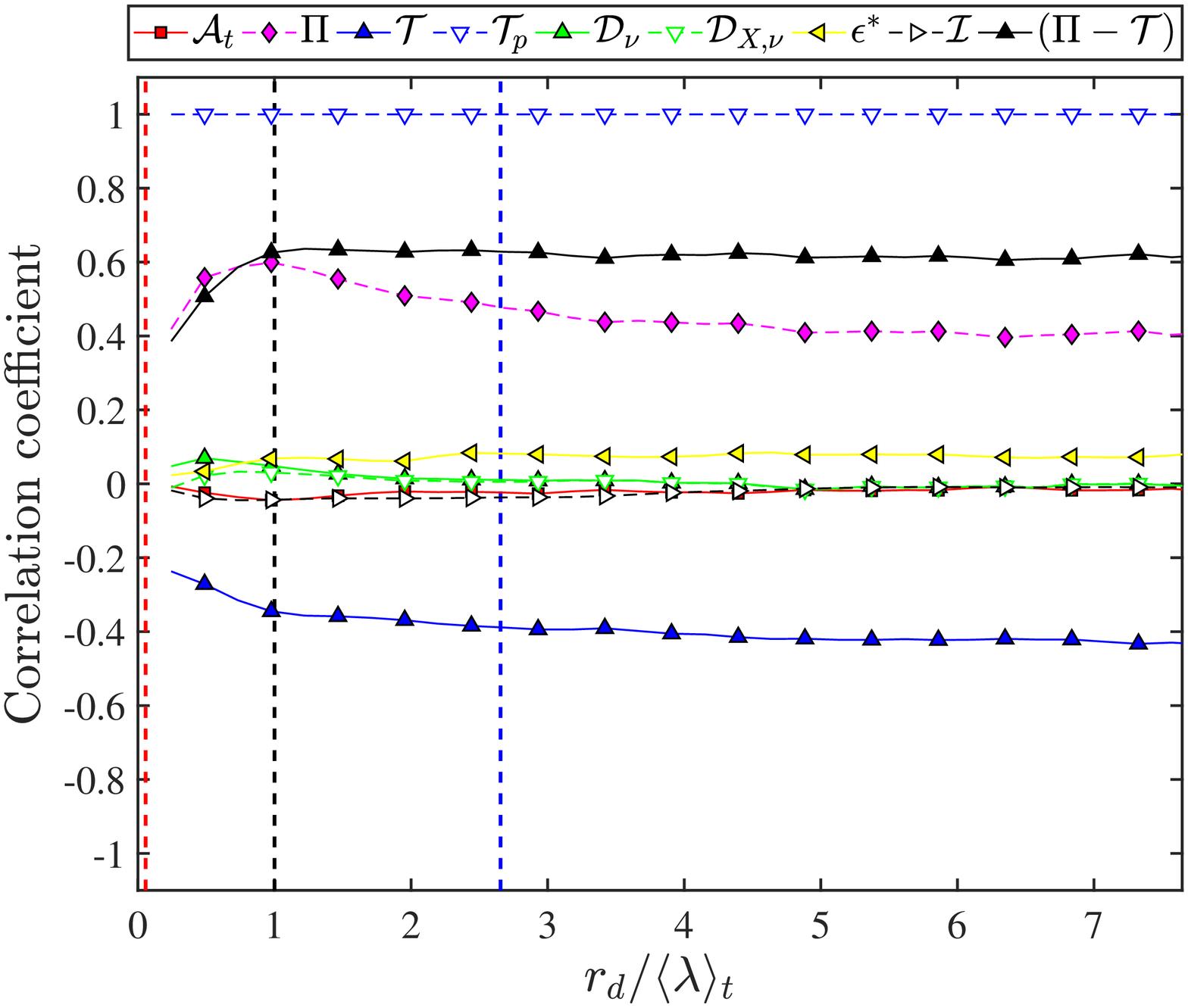}
        \end{minipage}
}
\subfigure{
        \begin{minipage}{0.5\linewidth}
                \centerline{(a2)}
                \includegraphics[clip,width=\linewidth]
                {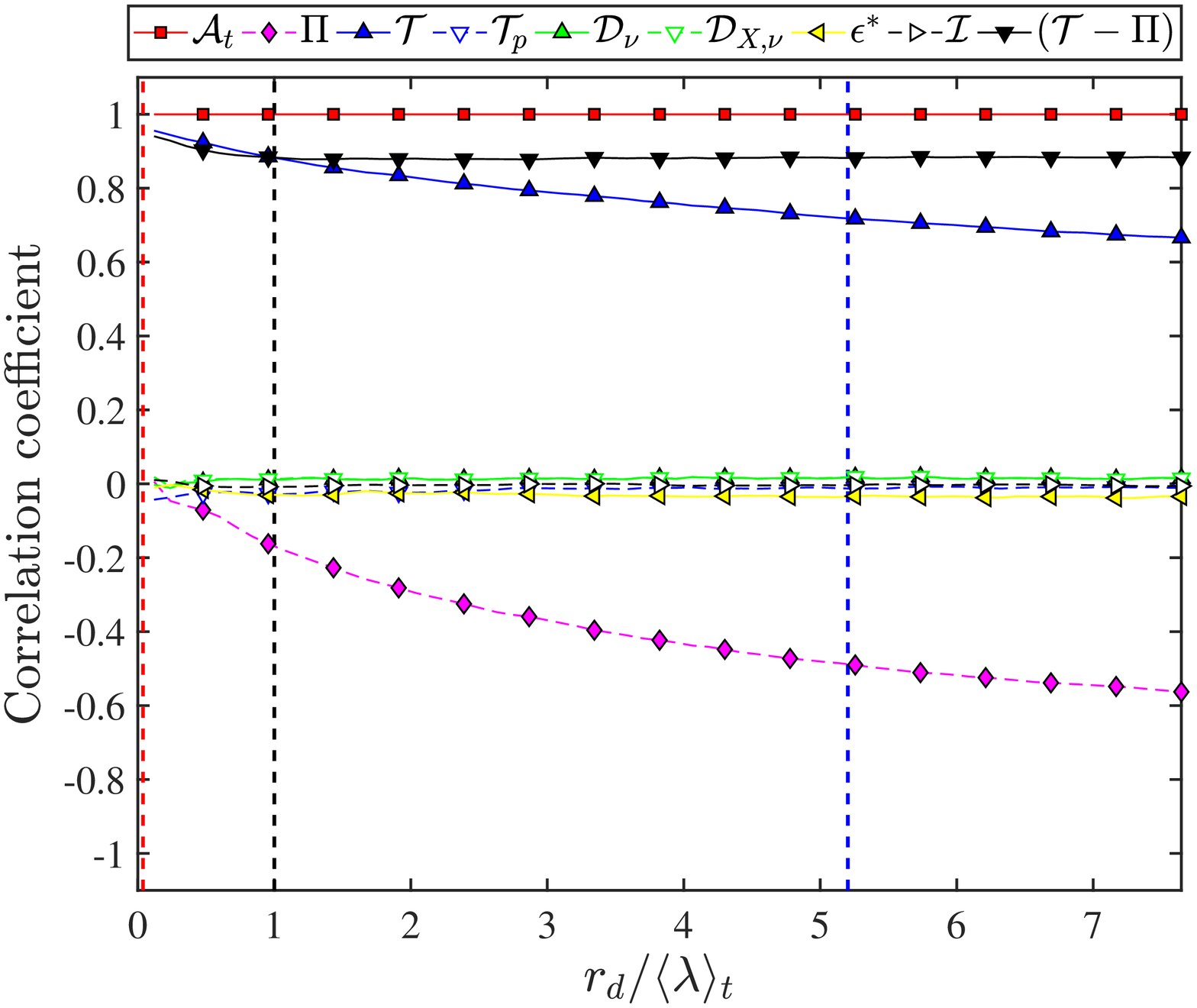}
        \end{minipage}
        \begin{minipage}{0.5\linewidth}
                \centerline{(b2)}
                \includegraphics[clip,width=\linewidth]
                {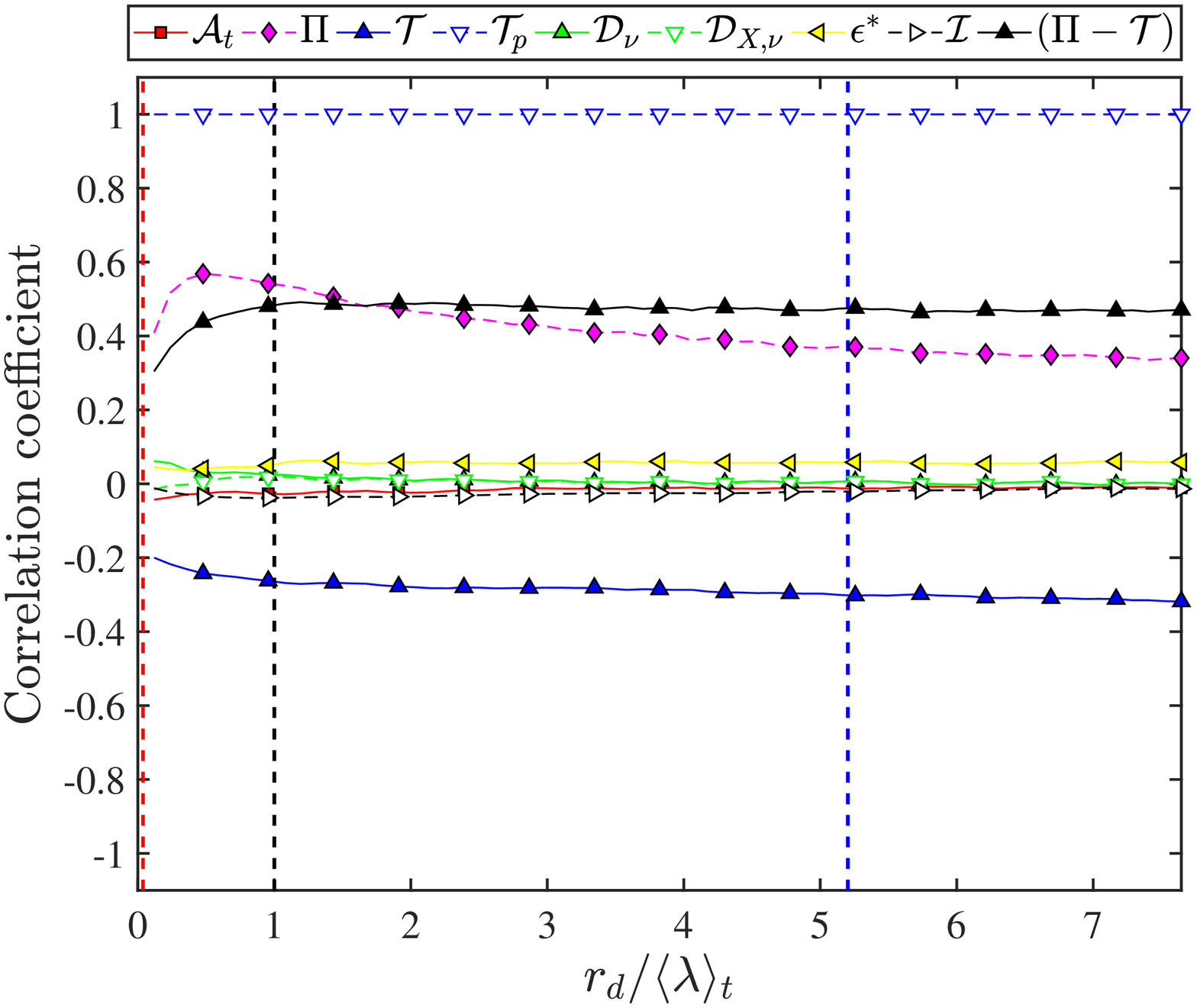}
        \end{minipage}
}
\caption{
Correlation coefficients $corr (Q_{1}, Q_{2})$ (defined in the second
paragraph of subsection 4.1) for the nine $Q_{2}$ terms indicated
above the plots and (a1) $Q_{1} = \mathcal{A}_{t}$ and $\langle
Re_{\lambda} \rangle_t = 80.9$, (a2) $Q_{1} = \mathcal{A}_{t}$ and
$\langle Re_{\lambda} \rangle_t = 178$, (b1) $Q_{1} = \mathcal{T}_p$
and $\langle Re_{\lambda} \rangle_t = 80.9$, (b2) $Q_{1} =
\mathcal{T}_p$ and $\langle Re_{\lambda} \rangle_t = 178$.  These
correlation coefficients are calculated by sampling over space ${\bm
  X}$, time $t$ and orientations $\hat{\bm r}$ for a given $r_d$ such
that ${\bm r} = r_{d} \hat{\bm r}$. They are plotted as functions
$r_d$ and normalised by $\langle \lambda \rangle_{t}$.
The red, black and blue dashed vertical straight lines indicate the
time-average Kolmogorov scale $\langle \eta \rangle_t$, Taylor micro
scale $\langle \lambda \rangle_t$, and integral scale $\langle L
\rangle_t$, respectively.}
\label{fig:cor_coeff_khmh}
\end{figure}

\begin{figure}
        \begin{minipage}{0.5\linewidth}
                \centerline{(a)}
                \includegraphics[clip,width=\linewidth]
                {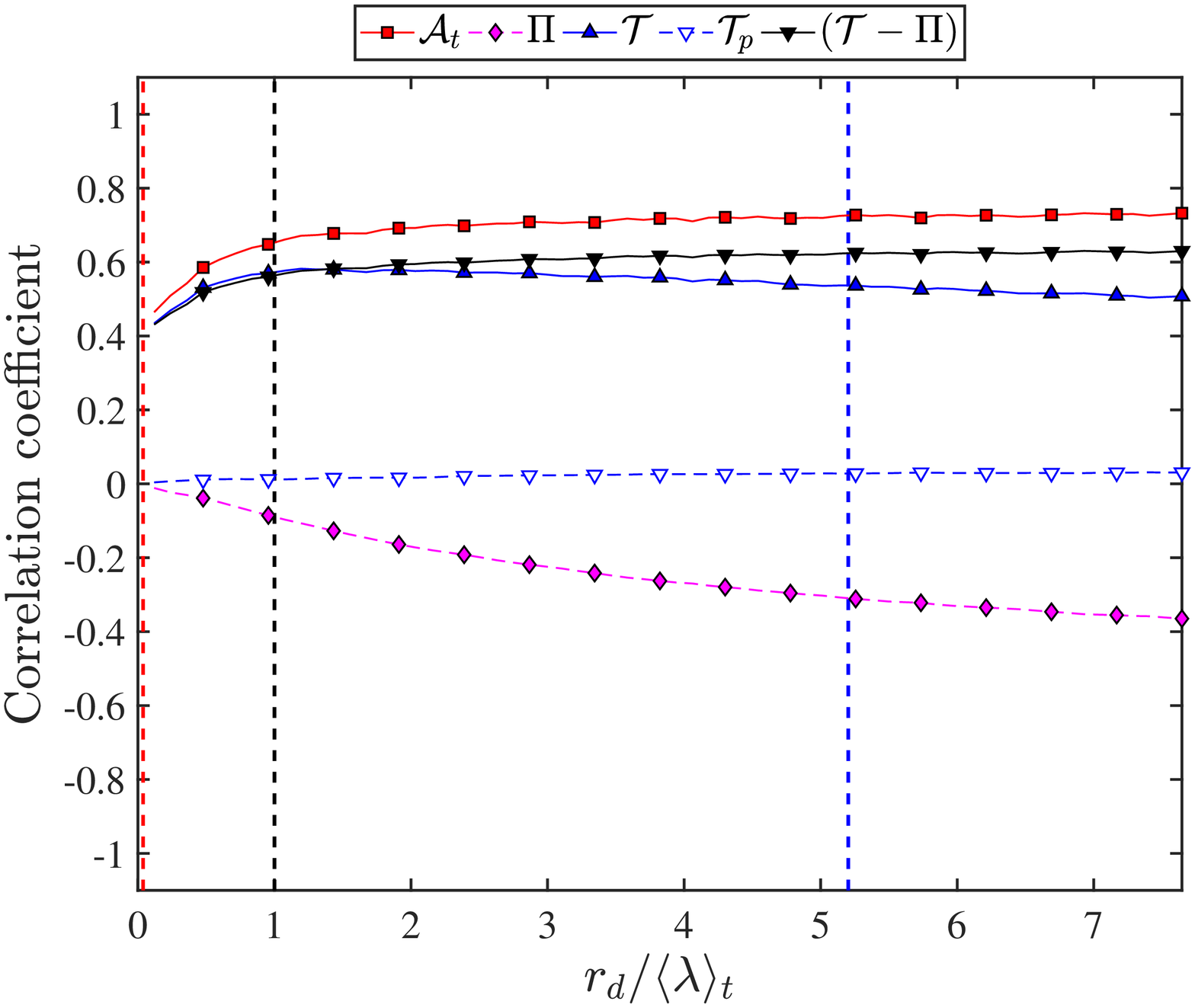}
        \end{minipage}
        \begin{minipage}{0.5\linewidth}
                \centerline{(b)}
                \includegraphics[clip,width=\linewidth]
                {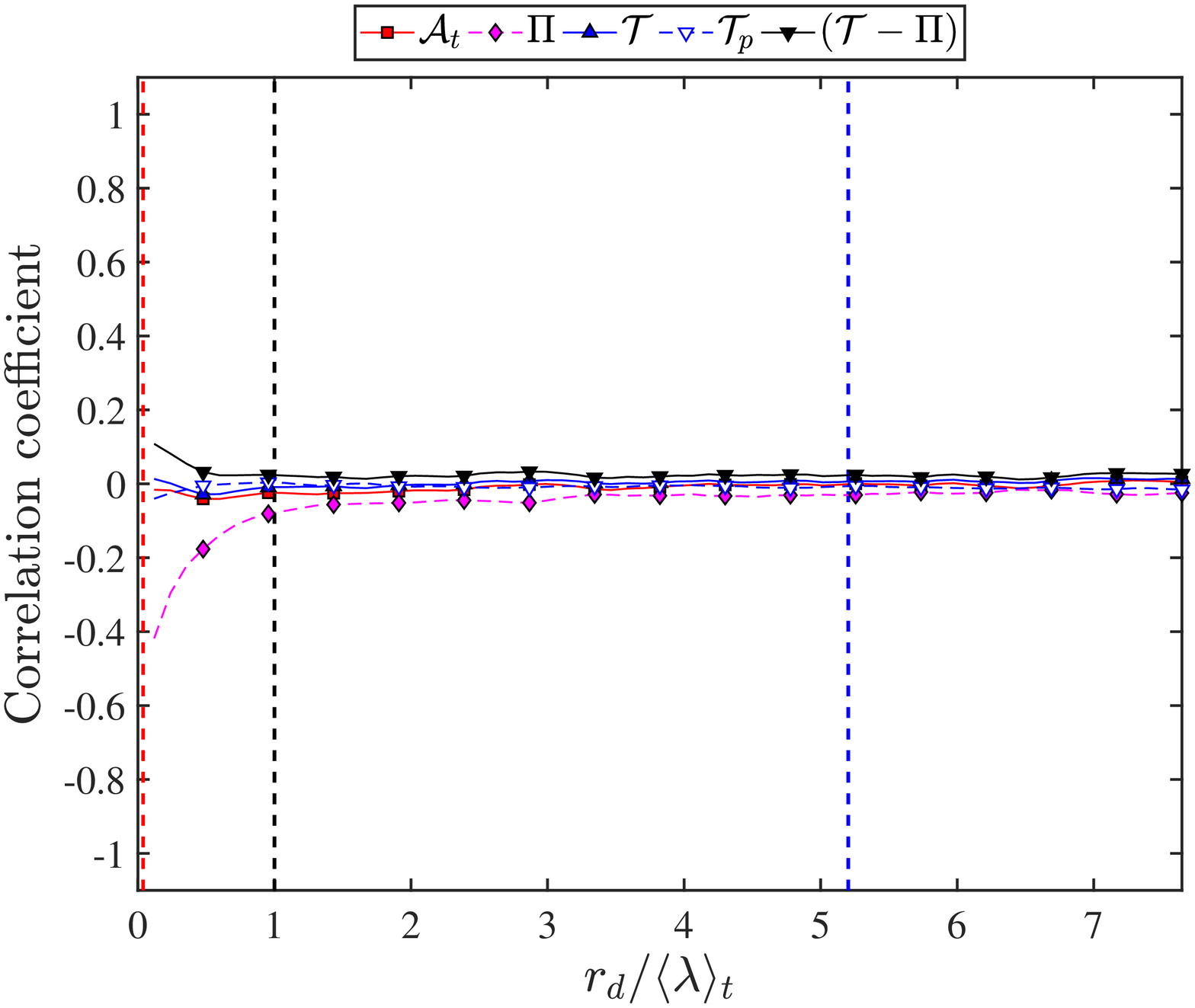}
        \end{minipage}
        \caption{(a) Correlation coefficients $corr (Q_{1}, Q_{2})$
          (defined in the second paragraph of subsection 4.1) for
          $Q_{1} = \cos \theta$ and $Q_{2}=$ $\mathcal{A}_t$, $\Pi$,
          $\mathcal{T}$, $\mathcal{T}_p$ and
          $\mathcal{T}-\mathcal{\Pi}$. (b) Correlation coefficients
          $corr (Q_{1}, Q_{2})$ for $Q_{1} = |\delta {\bm u} | |\delta
          \frac{\partial}{\partial t}{\bm u}|$ and $Q_{2}=$
          $\mathcal{A}_t$, $\Pi$, $\mathcal{T}$, $\mathcal{T}_p$ and
          $\mathcal{T}-\mathcal{\Pi}$. The correlation coefficients in
          these two plots are calculated and plotted as explained in
          the caption of figure~\ref{fig:cor_coeff_khmh}, and the red,
          black and blue dashed vertical straight lines are the same
          as in figure~\ref{fig:cor_coeff_khmh}. $\langle Re_{\lambda}
          \rangle_t = 178$ but these plots are very similar for
          $\langle Re_{\lambda} \rangle_t = 80.9$.}
\label{fig:cor_coeff_ang_du_ddudt}
\end{figure}

\begin{figure}
        \begin{minipage}{0.5\linewidth}
                \centerline{(a)}
                \includegraphics[clip,width=\linewidth]
                {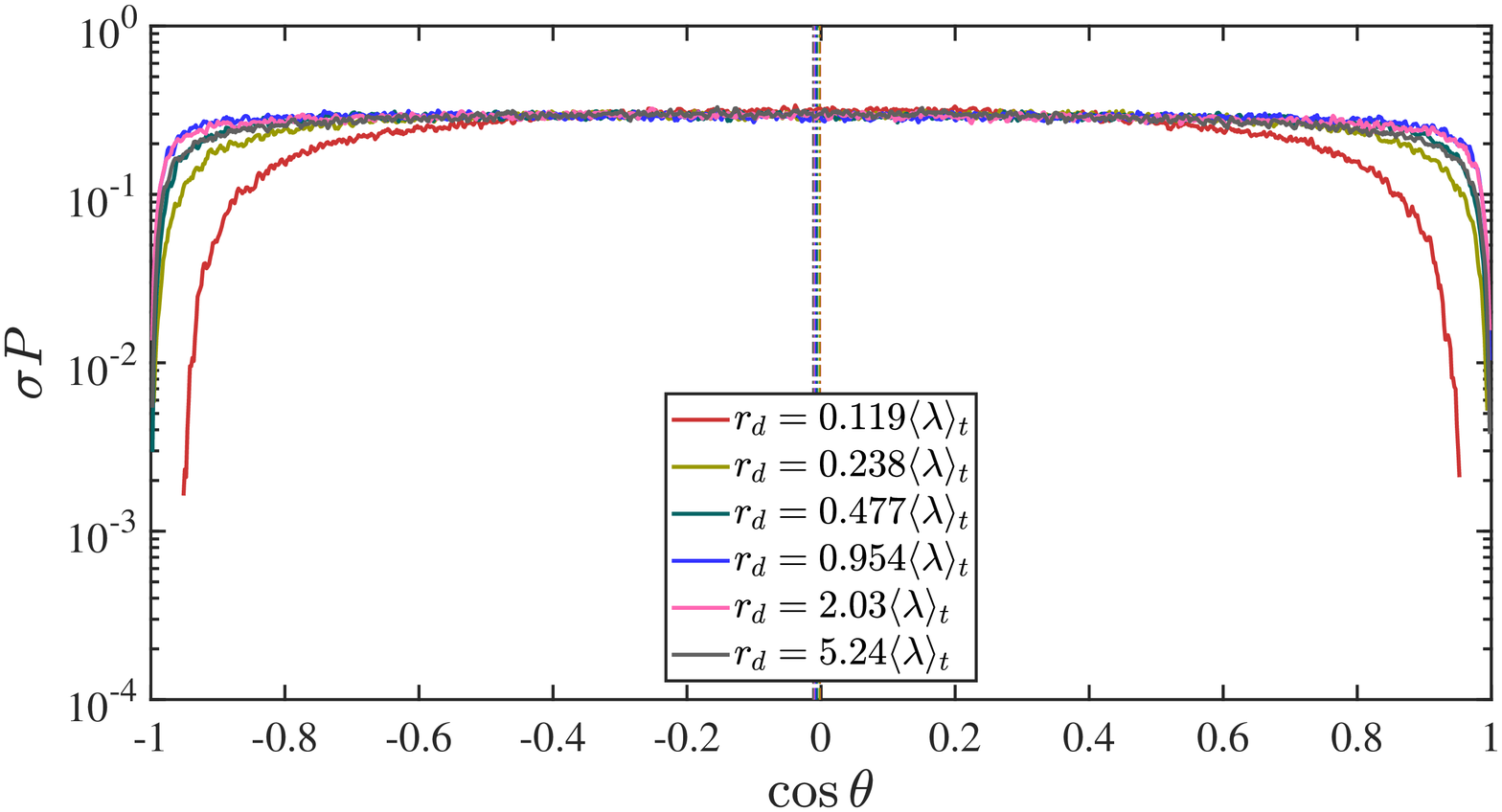}
        \end{minipage}
        \begin{minipage}{0.5\linewidth}
                \centerline{(b)}
                \includegraphics[clip,width=\linewidth]
                {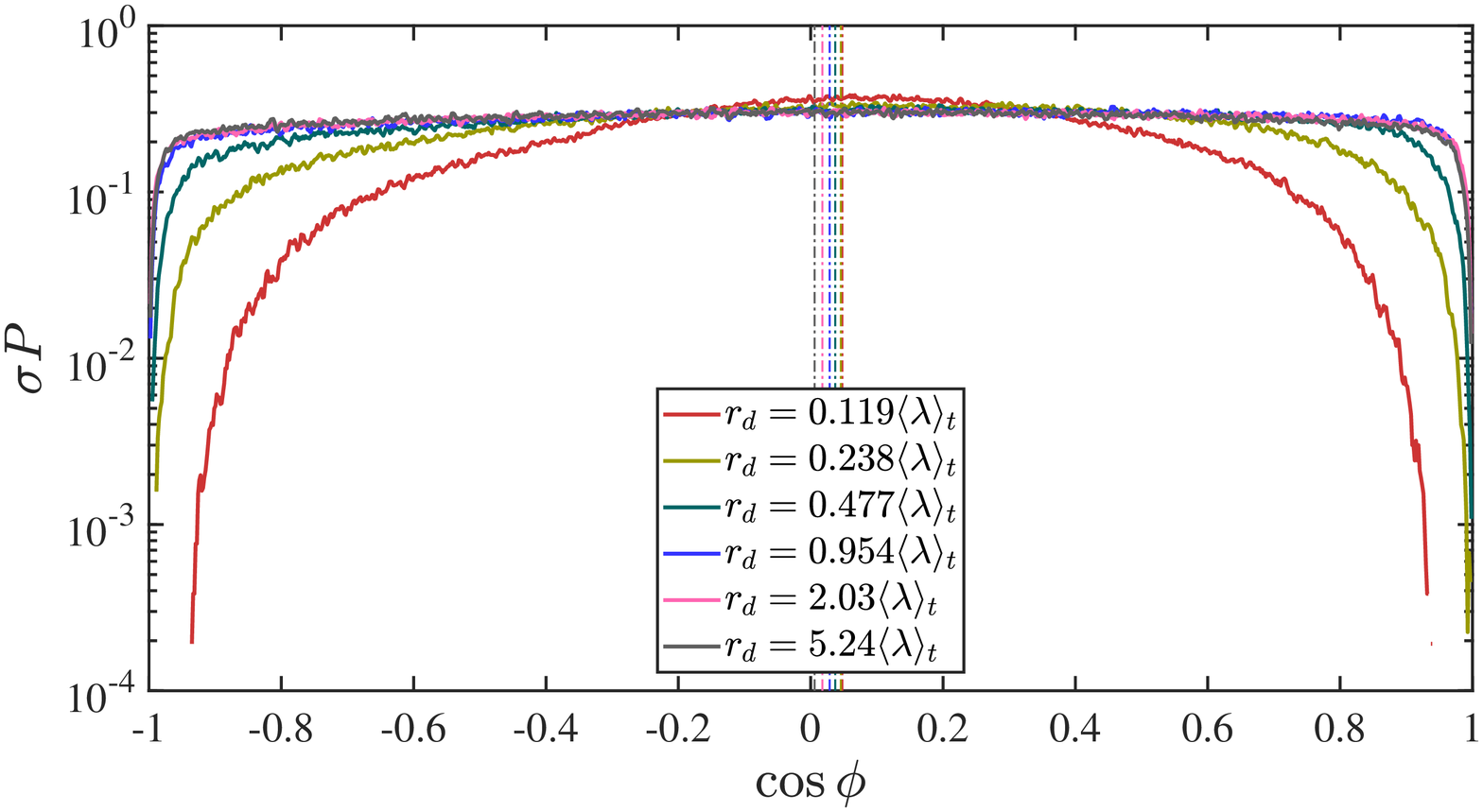}
        \end{minipage}
\caption{PDFs of cosines of angles obtained from sampling over ${\bm
    X}$, $t$ and $\hat{\bm r}$ for a given $r_d = |{\bm r}|$.
  Different curves correspond to different values of $r_{d}$ as shown
  in the insert. (a) PDFs of $\cos\theta$ multiplied by the standard
  deviation of $\cos\theta$. The vertical line indicates the average
  $\langle \cos\theta \rangle_{all}$ (average over ${\bm X}$, $t$ and
  $\hat{\bm r}$) which is effectively $0$ for all $r_d$. 
  (b) PDFs of $\cos\phi$ multiplied by the standard deviation of 
  $\cos\phi$. The vertical lines indicate the average 
  $\langle \cos\phi \rangle_{all}$ for different values 
  of $r_d$ from $0.119 \langle
  \lambda\rangle_{t}$ to $5.24 \langle \lambda\rangle_{t}$. The
  average of $\cos\phi$ is very close to $0$ for $r_{d} \le
  5.24\langle \lambda \rangle_{t}$ ($\approx \langle L \rangle_{t}$);
  in the range below $5.24\langle \lambda \rangle_{t}$, it increases
  monotonically with decreasing $r_d$ till it reaches $\approx 0.05$
  at $r_{d}=0.119\langle \lambda\rangle_{t}$. $\langle Re_{\lambda}
  \rangle_t = 178$ but these plots are very similar for $\langle
  Re_{\lambda} \rangle_t = 80.9$.}
\label{fig:pdf_ang_du_dpg}
\end{figure}

\begin{figure}
\centering
\subfigure{
        \begin{minipage}{0.5\linewidth}
                \centerline{(a1)}
                \includegraphics[clip,width=\linewidth]
                {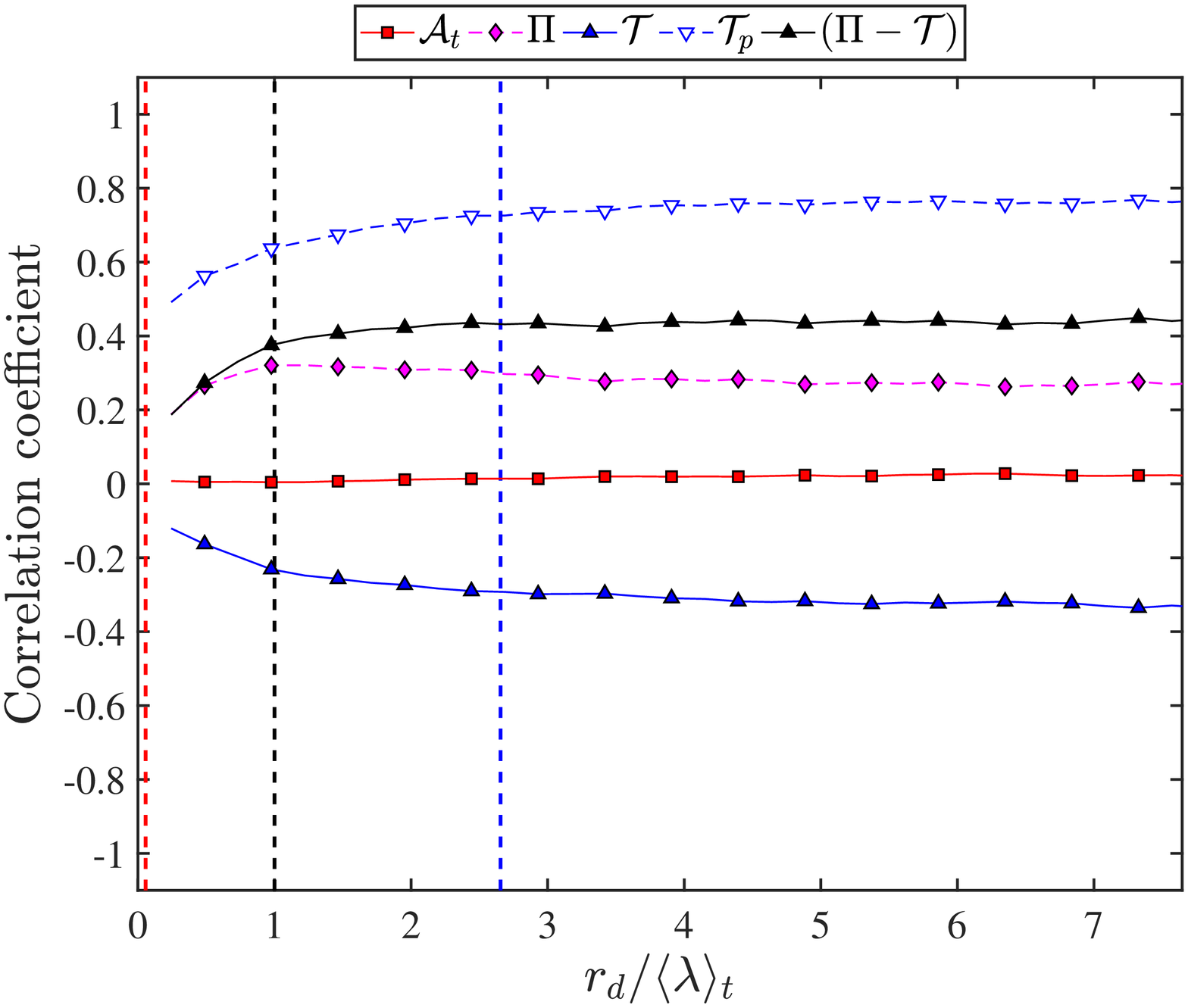}
        \end{minipage}
        \begin{minipage}{0.5\linewidth}
                \centerline{(b1)}
                \includegraphics[clip,width=\linewidth]
                {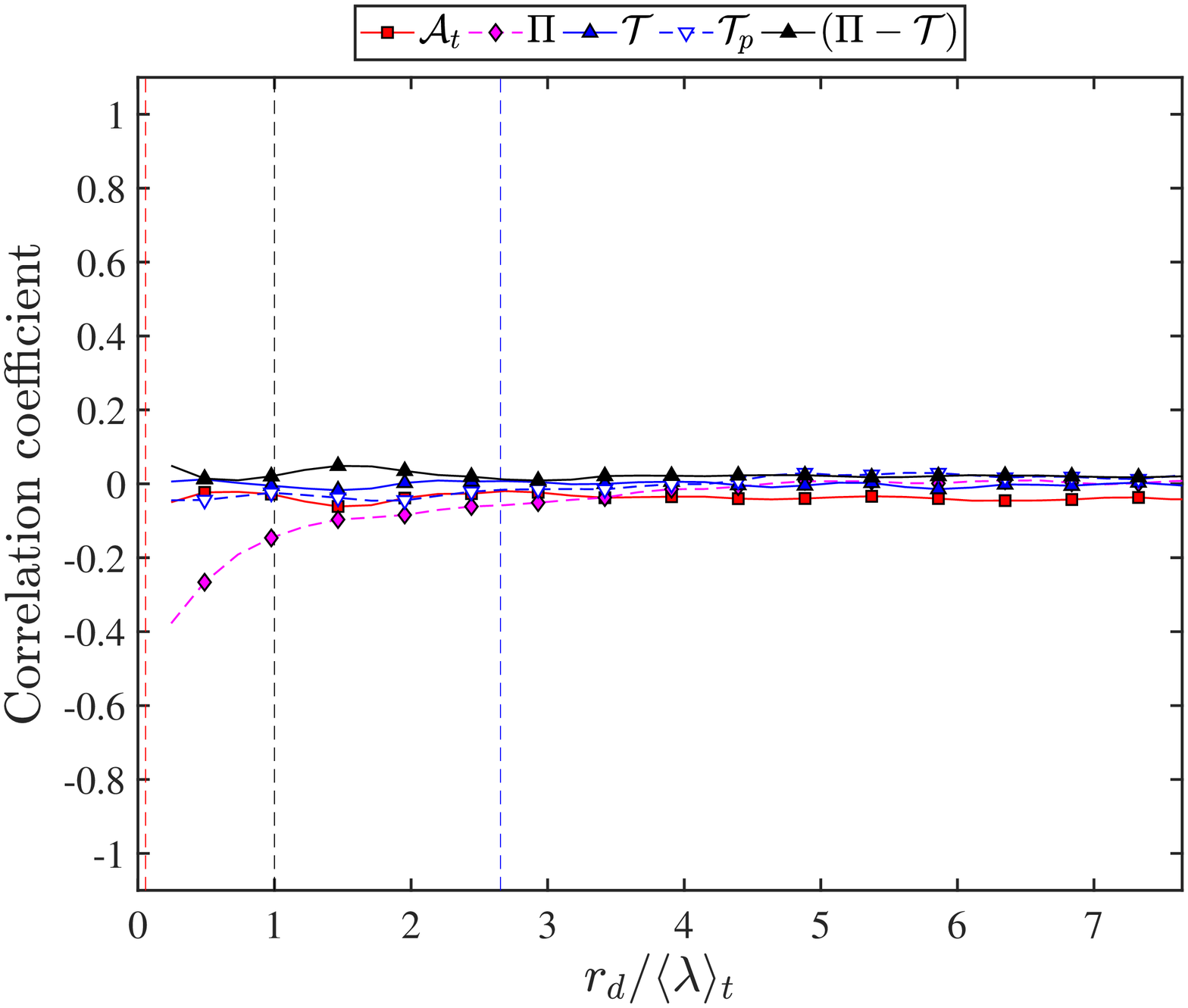}
        \end{minipage}
}
\subfigure{
        \begin{minipage}{0.5\linewidth}
                \centerline{(a2)}
                \includegraphics[clip,width=\linewidth]
                {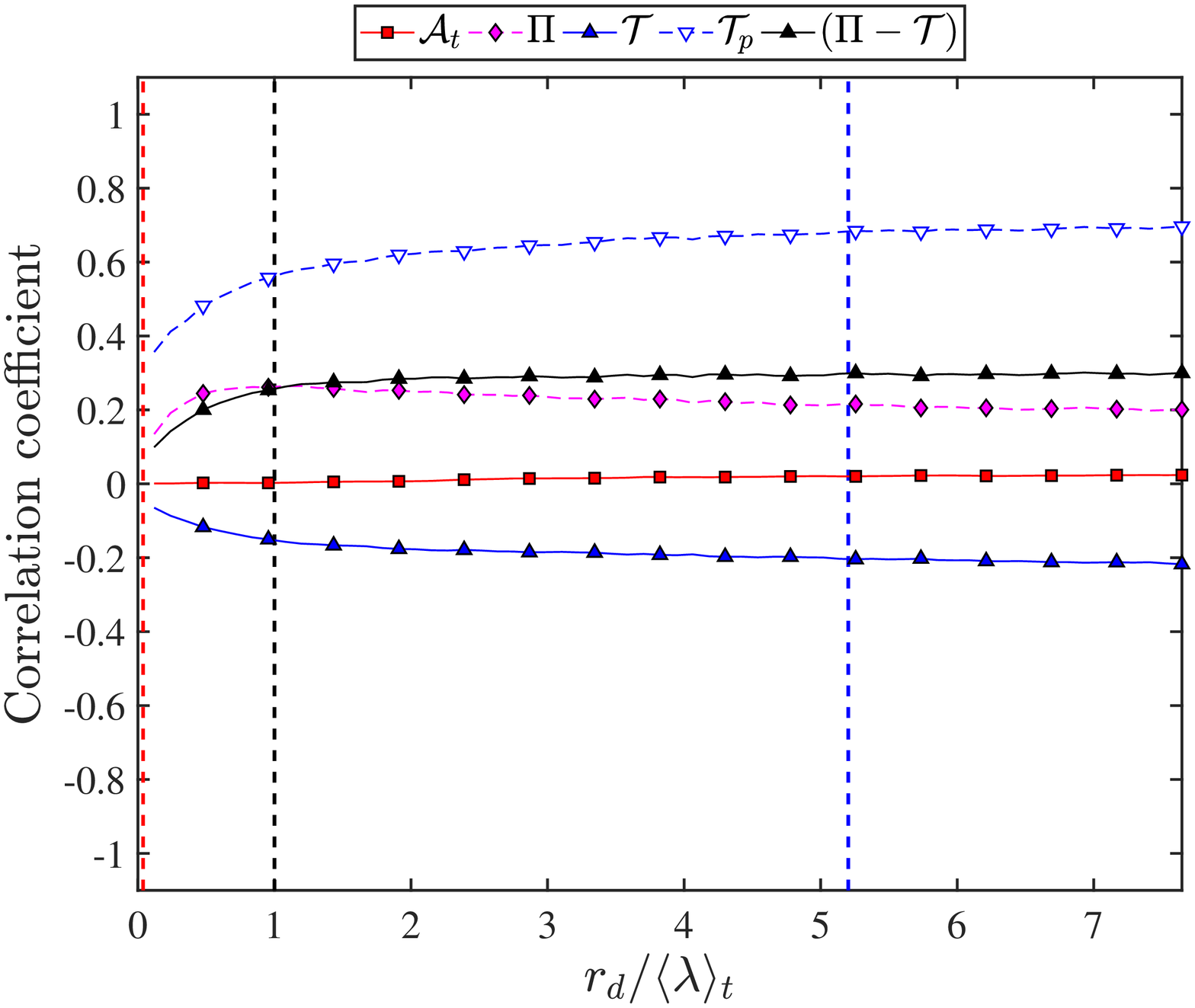}
        \end{minipage}
        \begin{minipage}{0.5\linewidth}
                \centerline{(b2)}
                \includegraphics[clip,width=\linewidth]
                {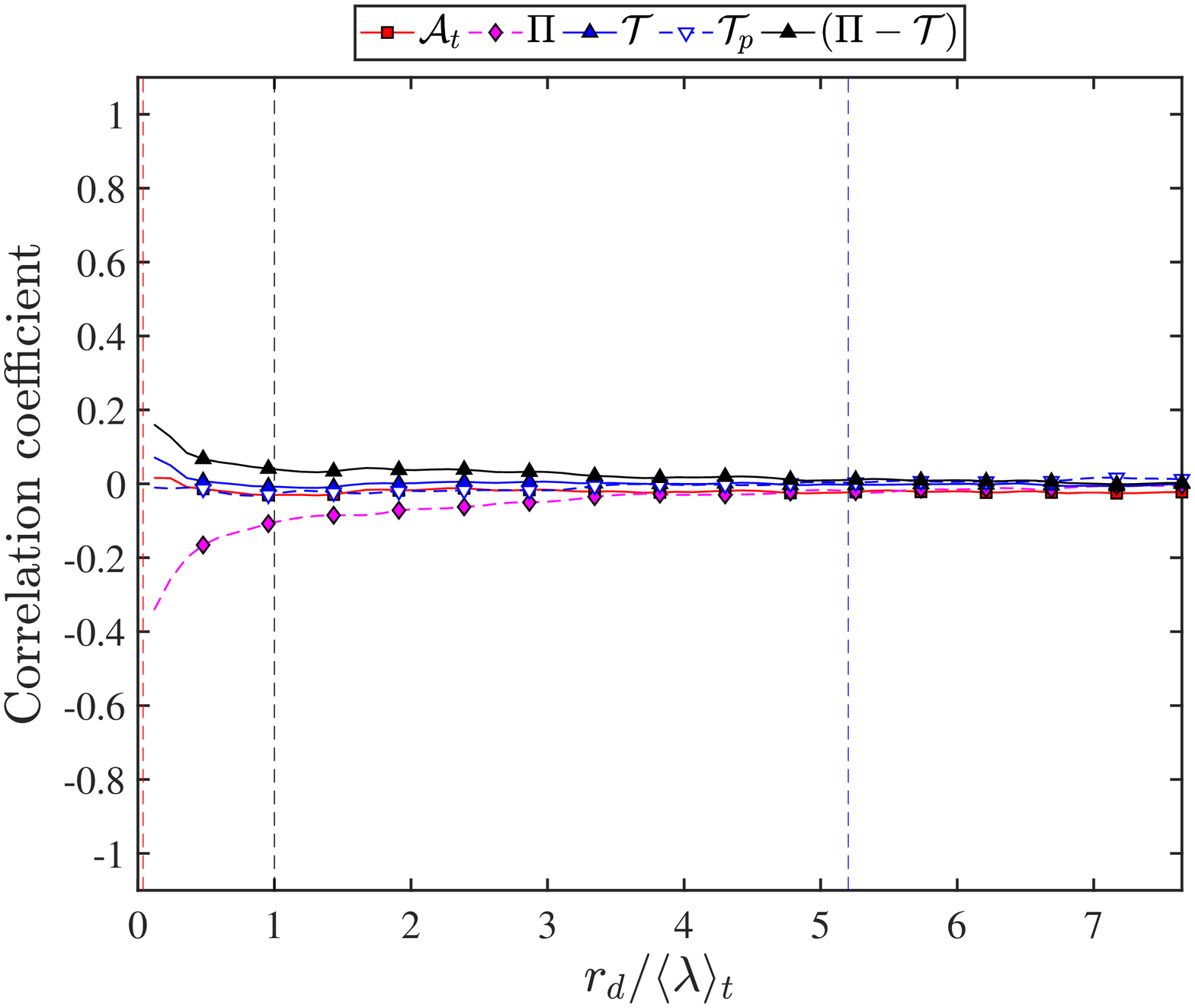}
        \end{minipage}       
}
        \caption{ (a1-a2) Correlation coefficients $corr (Q_{1},
          Q_{2})$ (defined in the second paragraph of subsection 4.1)
          for $Q_{1} = \cos \phi$ and $Q_{2}=$ $\mathcal{A}_t$, $\Pi$,
          $\mathcal{T}$, $\mathcal{T}_p$ and
          $\mathcal{\Pi}-\mathcal{T}$: (a1) for $\langle Re_{\lambda}
          \rangle_t = 80.9$, (a2) for $\langle Re_{\lambda} \rangle_t
          = 178$.  (b1-b2) Correlation coefficients $corr (Q_{1},
          Q_{2})$ for $Q_{1} = |\delta {\bm u} | |\delta {\bm f_{p}}|$
          (where ${\bm f_{p}}=-{\bm \nabla} p$) and $Q_{2}=$
          $\mathcal{A}_t$, $\Pi$, $\mathcal{T}$, $\mathcal{T}_p$ and
          $\mathcal{\Pi}-\mathcal{T}$; (b1) for $\langle Re_{\lambda}
          \rangle_t = 80.9$, (b2) for $\langle Re_{\lambda} \rangle_t
          = 178$. The correlation coefficients in these plots are
          calculated and plotted as explained in the caption of
          figure~\ref{fig:cor_coeff_khmh}, and the red, black and blue
          dashed vertical straight lines are the same as in
          figure~\ref{fig:cor_coeff_khmh}.}
\label{fig:cor_coeff_ang_du_dpg}
\end{figure}

The observations concerning fluctuation magnitudes are quantified in
terms of standard deviations of all orientation-averaged KHMH terms,
namely the standard deviations of $\mathcal{A}_{t}^{a}$,
$\mathcal{T}^a$, $\mathcal{\Pi}^a$, $\mathcal{T}_{p}^a$,
$\mathcal{I}^a$, $\mathcal{D}_{X,\nu}^a$, $\mathcal{D}_{\nu}^a$ and
$\mathcal{\epsilon}^{*a}$. These standard deviations are calculated by
averaging over position ${\bm X}$ and time $t$ and are presented in
two stacked column charts in figure~\ref{fig:deviations} for $r_d /
\langle \lambda \rangle_t$ between $0$ and $7.65$ for the $\langle
Re_{\lambda} \rangle_t = 178$ case. In figure~\ref{fig:deviations}(a)
the standard deviations are normalised by the sum of all standard
deviations at the value of $r_d / \langle \lambda \rangle_t$ where
this sum is maximal. This shows that all fluctuations increase with
decreasing $r_d$ and reach their most intense level at $r_d / \langle
\lambda \rangle_t \approx 1$, the value of $r_d$ where $-\langle
\Pi^{a}\rangle$ is maximal (see figure~\ref{fig:khmh_balance}). In
figure~\ref{fig:deviations}(b) the standard deviations at a given
$r_d$ are normalised by the sum of all standard deviations at that
$r_d$. This makes it clear that at any given scale $r_d$ the most
intense fluctuations are those of $\mathcal{A}_{t}^{a}$ and
$\mathcal{T}^a$ followed closely by the fluctuations of
$\mathcal{\Pi}^a$ and $\mathcal{T}_{p}^a$. These fluctuations are much
more intense than those of $\mathcal{I}^a$, $\mathcal{D}_{X,\nu}^a$,
$\mathcal{D}_{\nu}^a$ and $\mathcal{\epsilon}^{*a}$ except in the deep
dissipation range where the fluctuations of $\mathcal{D}_{\nu}^a$ and
$\mathcal{\epsilon}^{*a}$ increase in relative terms. All these
observations are the same for $\langle Re_{\lambda} \rangle_t = 80.9$
with the additional observation that the fluctuations of
$\mathcal{A}_{t}^{a}$ and $\mathcal{T}^a$ are appreciably more intense
relative to the other KHMH terms at $\langle Re_{\lambda} \rangle_t =
178$ than at $\langle Re_{\lambda} \rangle_t = 80.9$. 

\subsection{Correlations between different fluctuations}
Having found that the fluctuating and intermittent nature of the
scale-by-scale energy budget is not limited to the interscale energy
transfer rate but that the pressure-velocity term is equally
fluctuating and the time-derivative and turbulent transport terms are
fluctuating even more intensely, we now move to the correlations
between different fluctuating terms which we detected qualitatively in
figure~\ref{fig:timeevol_KHMH}. 

To quantify correlations between pairs of fluctuating terms $Q_1$ and
$Q_2$ in the KHMH equation, we use the correlation coefficient
$corr(Q_{1},Q_{2}) = \langle (Q_{1}-\langle Q_{1}\rangle_{all})
(Q_{2}-\langle Q_{2}\rangle_{all}) \rangle_{all} /
(\sigma_{Q_{1}}\sigma_{Q_{2}})$ where the averages $\langle
\cdot \rangle_{all}$ and the standard deviations $\sigma_{Q_{1}}$ of
$Q_{1}$ and $\sigma_{Q_{2}}$ of $Q_{2}$ are calculated by sampling
over space ${\bm X}$, time $t$ and all orientations $\hat{\bm r}$ for
a given $r_d$ such that ${\bm r} = r_{d} \hat{\bm r}$. We plot these
correlation coefficients in Figure~\ref{fig:cor_coeff_khmh} as
functions of $r_{d}$. The most notable correlations at length-scales
$r_d$ larger than $\langle \lambda \rangle_{t}$ are between:

(a) $Q_{1} = \mathcal{A}_{t}$ and $Q_{2}= \mathcal{T}-\mathcal{\Pi}$;
$Q_{1} = \mathcal{A}_{t}$ and $Q_{2} = \mathcal{T}$; $Q_{1} =
\mathcal{A}_{t}$ and $Q_{2} = -\mathcal{\Pi}$;

(b) $Q_{1} = \mathcal{T}_p$ and $Q_{2} = \mathcal{\Pi} - \mathcal{T}$;
$Q_{1} = \mathcal{T}_p$ and $Q_{2} = \mathcal{\Pi}$; $Q_{1} =
\mathcal{T}_p$ and $Q_{2} = -\mathcal{T}$.

The strongest correlation is between $\mathcal{A}_{t}$ and
$\mathcal{T}-\mathcal{\Pi}$ (see
figure~\ref{fig:cor_coeff_khmh}(a1-a2)), 
equally so over the entire range of scales 
$r_{d} > \langle \lambda \rangle_{t}$. This range
includes the embryonic approximately inertial range between $\langle
\lambda \rangle_{t}$ and $\langle L \rangle_{t}$ as well as scales
larger than $\langle L \rangle_{t}$. This strong correlation is a
reflection of the sweeping effect \citep[see][]{tsinober_2014_an}
whereby the local acceleration ${\partial \over \partial t} \bm{u}$
and the convective acceleration $\bm{u}\cdot \bm{\nabla} \bm{u}$
have a tendency to approximately cancel each other. The direct
mathematical consequence on the KHMH equation is that
$\mathcal{A}_{t}$ and $\mathcal{T}-\mathcal{\Pi}$ would tend to cancel
each other too, and the strong correlation seen in
figure~\ref{fig:cor_coeff_khmh}(a1-a2) between these two terms is a
direct confirmation of this cancellation tendency. Note that this
correlation seems to strengthen with increasing $\langle Re_{\lambda}
\rangle_t$.

The correlations between $\mathcal{A}_{t}$ and $\mathcal{T}$ on the
one hand, and between $\mathcal{A}_{t}$ and $-\mathcal{\Pi}$ on the
other are weaker. However, the correlation between $\mathcal{A}_{t}$
and $\mathcal{T}$ strengthens whereas the correlation between
$\mathcal{A}_{t}$ and $-\mathcal{\Pi}$ weakens and tends to $0$ with
decreasing $r_d$, particularly in the range $r_{d} \le \langle L
\rangle_{t}$, see figure~\ref{fig:cor_coeff_khmh}(a1-a2). It is
significant that the interscale energy transfer term decorrelates from
the sweeping effect as the scale decreases within the inertial range.
Note also that the correlation between $\mathcal{A}_{t}$ and
$\mathcal{T}$ is slightly stronger for our higher $\langle
Re_{\lambda} \rangle_t$ values.

\subsection{Geometrical statistics}
The sweeping effect goes some considerable way in explaining the set
(a) of correlations above. However more insight can be obtained with a
geometrical approach in terms of the angle $\theta ({\bm X}, {\bm r},
t)$ between the two-point vectors $\delta \bm{u}$ and $\delta
\frac{\partial}{\partial t}\bm{u}$. In terms of this angle, the term
$\mathcal{A}_t$ reads
\begin{eqnarray}
\label{eq:trnsp_A_angle} 
4\mathcal{A}_t = |\delta \bm{u}| |\delta \frac{\partial}{\partial
    t} \bm{u}| \cos \theta
\:,
\end{eqnarray}
and its correlations with $|\delta \bm{u}| |\delta
\frac{\partial}{\partial t}\bm{u}|$ and $\cos \theta$ separately are
plotted in figure~\ref{fig:cor_coeff_ang_du_ddudt}. The correlation is
strong with $\cos \theta$ (figure~\ref{fig:cor_coeff_ang_du_ddudt}(a))
but inexistent with $|\delta \bm{u}| |\delta \frac{\partial}{\partial
  t}{\bm u}|$ (figure~\ref{fig:cor_coeff_ang_du_ddudt}(b)). Given that
$\langle \mathcal{A}_t\rangle_{all} =0$ and that $\langle \cos \theta
\rangle_{all}$ is also effectively zero for all $r_d$ (see
figure~\ref{fig:pdf_ang_du_dpg}(a)) we conclude that $|\delta \bm{u}|
|\delta \frac{\partial}{\partial t}{\bm u}|$ and $\cos\theta$ are
uncorrelated at all scales (a fact that we confirmed against our data
for all $r_d$) and that the sign of $\mathcal{A}_t$ is therefore
purely dictated by the sign of $\cos\theta$ independently from the value
of $|\delta {\bm u}| |\delta \frac{\partial}{\partial t}{\bm
  u}|$. This is reflected in the strong correlation between
$\mathcal{A}_t$ and $\cos\theta$ in
figure~\ref{fig:cor_coeff_ang_du_ddudt}(a). Note that the probability
of $\cos\theta$ is in fact equally distributed, with the exception of
perfect or near-perfect alignments or anti-alignments between $\delta
{\bm u}$ and $\delta \frac{\partial}{\partial t}{\bm u}$ which are
particularly unlikely (see figure~\ref{fig:pdf_ang_du_dpg}(a)).

Figure~\ref{fig:cor_coeff_ang_du_ddudt}(b) shows that $\mathcal{T}$
and $\mathcal{T}_{p}$ are also totally uncorrelated with $|\delta {\bm
  u}| |\delta \frac{\partial}{\partial t}{\bm u}|$ at all scales and
that the same holds for $-\mathcal{\Pi}$ and
$\mathcal{T}-\mathcal{\Pi}$ except for very small correlations with
$|\delta {\bm u}| |\delta \frac{\partial}{\partial t}{\bm u}|$ at
scales $r_d$ below $\langle L\rangle_{t}$.

Figure~\ref{fig:cor_coeff_ang_du_ddudt}(a) reveals that the set (a) of
correlations caused by the sweeping effect between $\mathcal{A}_{t}$
and each one of the terms $\mathcal{T}-\mathcal{\Pi}$, $\mathcal{T}$
and $-\mathcal{\Pi}$ translates into correlations between $\cos \theta$ 
and each one of these terms, albeit a little weaker. In
agreement with the observation made at the end of the previous
sub-section, the correlation between $\cos\theta$ and $\mathcal{T}$
strengthens whereas the correlation between $\cos\theta$ and
$-\mathcal{\Pi}$ weakens and tends to $0$ with decreasing $r_d$,
particularly in the range $\langle \lambda \rangle_{t} \le r_{d} \le
\langle L \rangle_{t}$, see figure~\ref{fig:cor_coeff_ang_du_ddudt}(a)
and compare with figure~\ref{fig:cor_coeff_khmh}(a). The interscale
energy transfer term decorrelates from the sweeping alignment as the
scale decreases within the inertial range and as the correlation
between the turbulent transport term and $\cos\theta$ strengthens.

Finally, note the zero correlation between $\mathcal{T}_p$ and
$\cos\theta$ in figure~\ref{fig:cor_coeff_ang_du_ddudt}(a), in
agreement with the zero-correlation between $\mathcal{T}_p$ and
$\mathcal{A}_t$ in figure~\ref{fig:cor_coeff_khmh}(a1-a2). However,
the pressure-velocity term $\mathcal{T}_p$ is central in the set (b)
of correlations (figure~\ref{fig:cor_coeff_khmh}(b)) mentioned in the
previous subsection. The very significant correlation between
$\mathcal{\Pi}-\mathcal{T}$ and $\mathcal{T}_p$ at all scales $r_d \ge
\langle \lambda \rangle_{t}$ reflects the intimate relation between
the convective non-linearity and the pressure gradient non-locality in
incompressible Navier-Stokes flow. It is also interesting that, whilst
this correlation remains constant (at about 0.6 for $\langle
Re_{\lambda} \rangle_t = 80.9$ and a little under 0.5 for $\langle
Re_{\lambda} \rangle_t = 178$) throughout the range $\langle \lambda
\rangle_t$ to $\langle L \rangle_t$ and beyond, the correlation
between $\mathcal{T}_p$ and $\mathcal{\Pi}$ increases whereas the
correlation between $\mathcal{T}_p$ and $\mathcal{T }$ decreases with
decreasing $r_d$ in this range. It appears that the correlation
between $\mathcal{\Pi}$ and $\mathcal{T}_p$ increases as
$\mathcal{\Pi}$ tends to assume its cascade role in an increasingly
exclusive way with diminishing interference from the sweeping effect.
This is clearer at our higher Reynolds number because, whilst the
correlation between $\mathcal{T}_p$ and $\mathcal{\Pi}$ is about the
same for the two Reynolds numbers, the correlation between
$\mathcal{T}_p$ and $-\mathcal{T }$ seems to decrease with increasing
$\langle Re_{\lambda} \rangle_t$.

For further insight we define the angle $\phi ({\bm X}, {\bm r}, t)$
between the two-point vectors $\delta {\bm u}$ and $-\delta {\bm
  \nabla}p$ and write
\begin{eqnarray}
\label{eq:trnsp_pp_angle} 
4\mathcal{T}_p \equiv \frac{2}{\rho} \delta u_i \bigg( - \delta
\frac{\partial p}{\partial x_i} \bigg) = \frac{2}{\rho} |\delta \bm{u}| 
|\delta {\bm \nabla} p| \cos \phi
\:.
\end{eqnarray}
Figure~\ref{fig:cor_coeff_ang_du_dpg} shows that $\mathcal{T}_p$ is
quite highly correlated with $\cos\phi$ but is not correlated with
$|\delta \bm{u}| |\delta {\bm \nabla} p|$. Furthermore, as we report
in figure~\ref{fig:pdf_ang_du_dpg}(b) and its caption and also in
figure~\ref{fig:mean_dot_cross}, $\langle \cos\phi\rangle_{all}$ is
not zero for scales $r_d$ smaller than $\langle L\rangle_t$. This
implies that there must be a correlation between $|\delta \bm{u}|
|\delta {\bm \nabla} p|$ and $\cos\phi$ at these scales. However, we
checked that this correlation is small which agrees with the
observation that $\langle \cos\phi\rangle_{all}$ is small too (see
caption of figure~\ref{fig:pdf_ang_du_dpg}). The sign of
$\mathcal{T}_p$ is therefore mainly determined by the sign of
$\cos\phi$ without significant interference from $|\delta \bm{u}|
|\delta {\bm \nabla} p|$, as can be seen in the strong correlation
between $\mathcal{T}_p$ and $\cos\phi$ in
figure~\ref{fig:cor_coeff_ang_du_dpg}(a). Note in
figure~\ref{fig:pdf_ang_du_dpg} the slight tendency for a preference
in angle $\phi$ as $r_d$ decreases in the range below
$\langle L\rangle_t$; this statistical preference is for slightly
positive values of $\cos\phi$.  One cannot help thinking that there
may be a relation with the slight tendency for a negative value of
$\Pi$ in figure~\ref{fig:pdf_TT}. There is definitely a significant
role of the fluctuating pressure in the interscale energy transfer
dynamics expressed by the correlations between $\mathcal{T}_p$ and
$\mathcal{\Pi}$.

Figure~\ref{fig:cor_coeff_ang_du_dpg}(b) shows that, like
$\mathcal{T}_p$, $\mathcal{T}$ is uncorrelated with $|\delta \bm{u}|
|\delta {\bm \nabla} p|$ at all scales too; $\mathcal{\Pi}$ and
$\mathcal{\Pi} -\mathcal{T}$ are also uncorrelated with 
$|\delta \bm{u}| |\delta {\bm \nabla} p|$ except at scales below $\langle
L\rangle_{t}$ and increasingly so with decreasing $r_d$. These small
correlations are similar to those between $|\delta {\bm u}| |\delta
\frac{\partial}{\partial t}{\bm u}|$ and either $\mathcal{\Pi}$ or
$\mathcal{\Pi} -\mathcal{T}$ in
figure~\ref{fig:cor_coeff_ang_du_ddudt}(b). They both seem to have
their origin in small-scale inertial and dissipation range
correlations between $|\delta {\bm u}|$ and $\mathcal{\Pi}$.

The KHMH terms $\mathcal{\Pi} -\mathcal{T}$, $-\mathcal{T}$ and
$\mathcal{\Pi}$ exhibit some significant correlation with $\cos\phi$
whereas $\mathcal{A}_{t}$, which is mostly affected by the sweeping,
does not (see figure~\ref{fig:cor_coeff_ang_du_dpg}(a)). The lack of
correlation between $\cos \phi$ and $\mathcal{A}_{t}$ and the
significant correlation between $\cos \phi$ and $\mathcal{\Pi}$ are
consistent with the lack of correlation between $\mathcal{T}_p$ and
$\mathcal{A}_{t}$ and the significant correlation between
$\mathcal{T}_p$ and $\mathcal{\Pi}$
(figure~\ref{fig:cor_coeff_khmh}). In the previous subsection we saw
how, as $r_d$ decreases within the inertial range, the sweeping effect
is progressively expressed exclusively by the strengthening
correlation between $\mathcal{T}$ and $\mathcal{A}_{t}$ as
$\mathcal{\Pi}$ progressively decorrelates from
$\mathcal{A}_{t}$. The slightly increasing positive correlation that
$\mathcal{\Pi}$ has with $\cos \phi$
(figure~\ref{fig:cor_coeff_ang_du_dpg}(a1-a2)) as $r_d$ decreases for
both $\langle Re_{\lambda} \rangle_t$ values reflect the same, albeit
stronger, effect between $\mathcal{\Pi}$ and $\mathcal{T}_p$
(figure~\ref{fig:cor_coeff_khmh}(b)).  A contributing factor to this
effect being weaker in figure~\ref{fig:cor_coeff_ang_du_dpg}(a1-a2)
than in figure~\ref{fig:cor_coeff_khmh}(b) is that both $-\mathcal{T}$
and $\mathcal{\Pi}-\mathcal{T}$ tend to reduce their correlation with
$\cos \phi$ as $\mathcal{T}$ assumes more of its sweeping effect role
by correlating further with $\mathcal{A}_{t}$ both with decreasing
$r_d$ below $\langle L \rangle_{t}$ and with increasing $\langle
Re_{\lambda} \rangle_t$. Even so, there seems to be an interference of
the pressure gradient fluctuations and their geometrical alignments
with the dynamics of interscale energy transfer.

\begin{figure}
        \begin{minipage}{0.5\linewidth}
                \centerline{(a)}
                \includegraphics[clip,width=\linewidth]
                {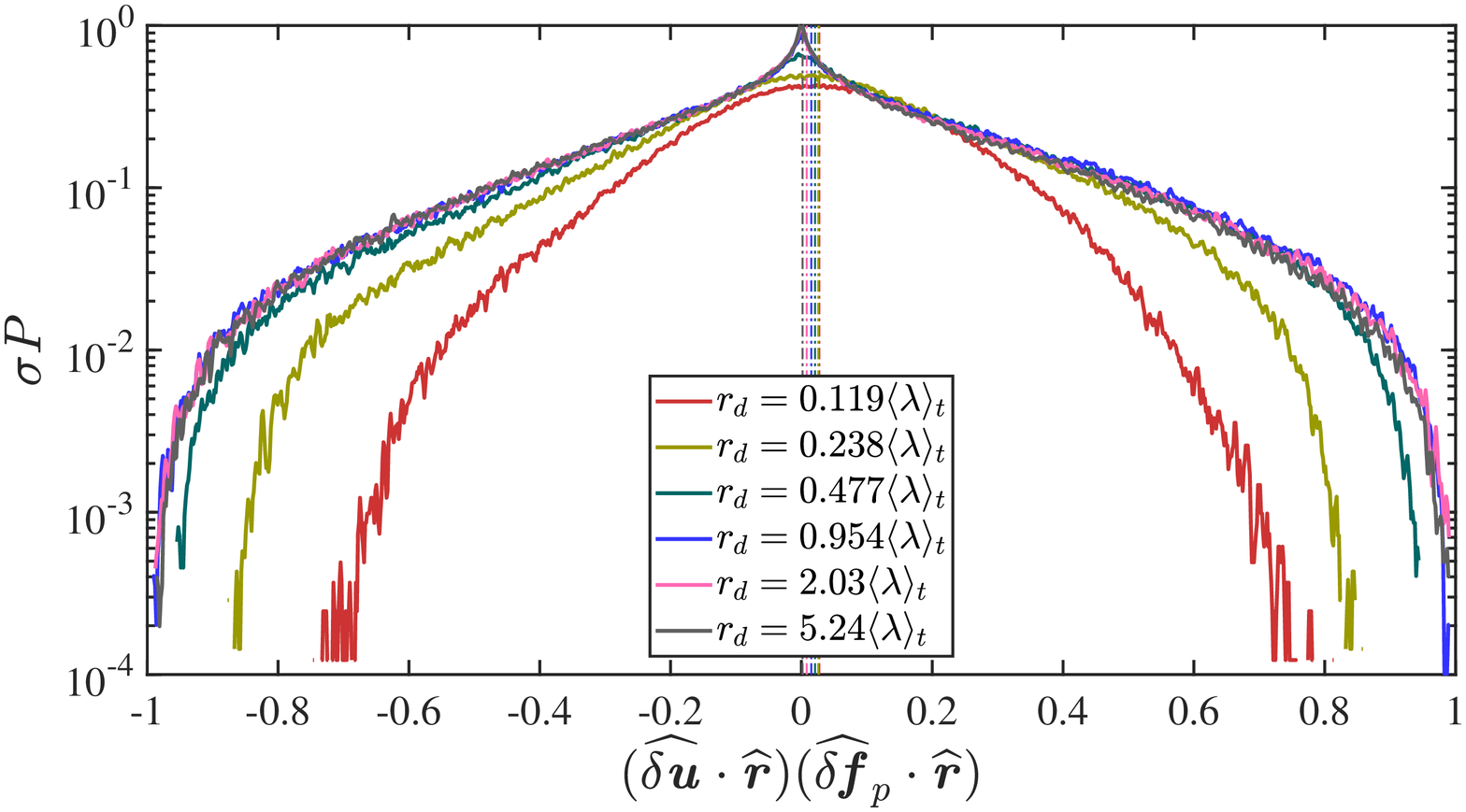}
        \end{minipage}
        \begin{minipage}{0.5\linewidth}
                \centerline{(b)}
                \includegraphics[clip,width=\linewidth]
                {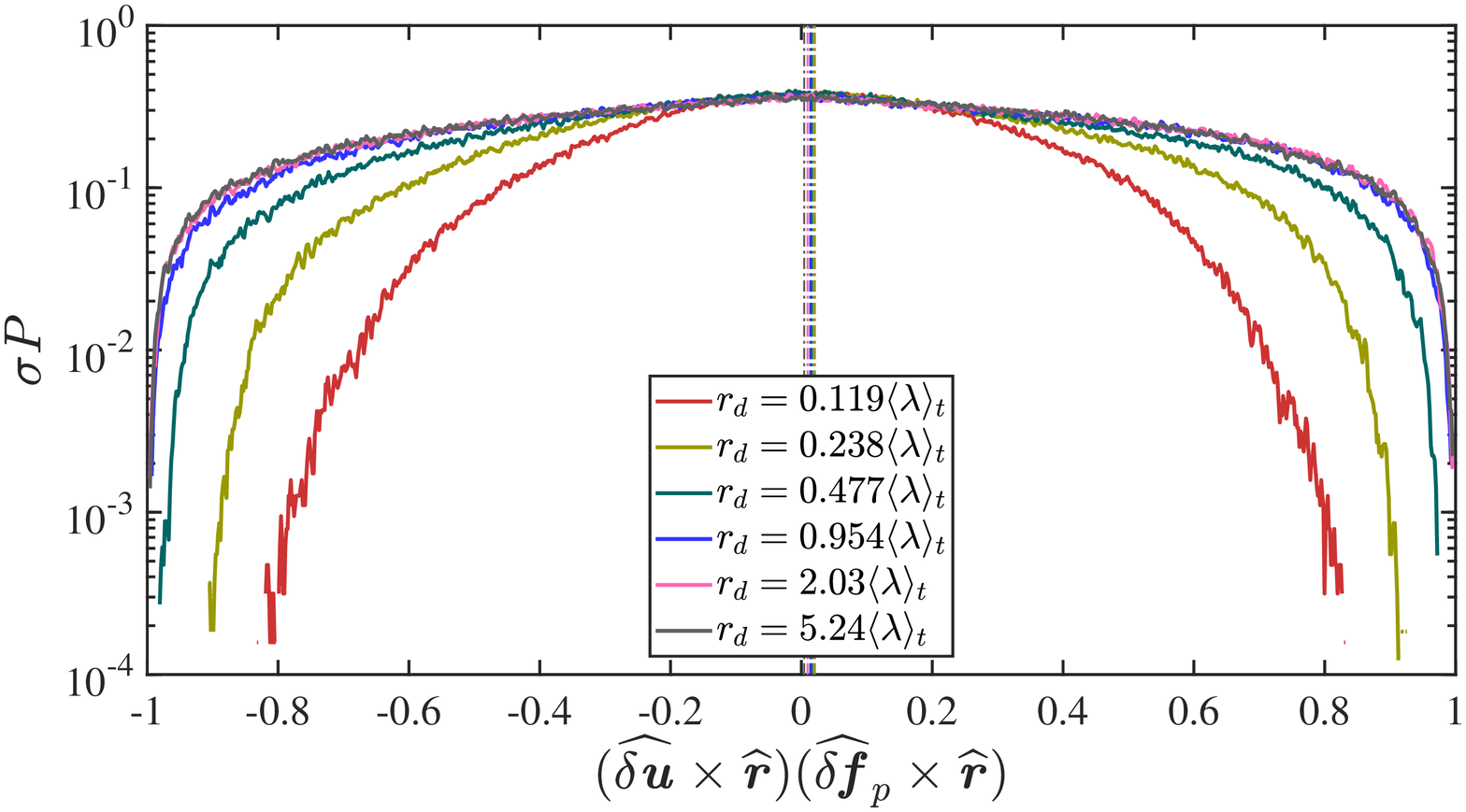}
        \end{minipage}
        \caption{PDFs of (a) $(\widehat{\delta {\bm u}}\cdot \hat{\bm
            r})(\widehat{\delta {\bm f}_{p}}\cdot \hat{\bm r})$ and
          (b) $(\widehat{\delta {\bm u}}\times \hat{\bm
            r})\cdot(\widehat{\delta {\bm f}_{p}}\times \hat{\bm r})$
          obtained from sampling over ${\bm X}$, $t$ and $\hat{\bm r}$
          for a given $r_d = |{\bm r}|$. The PDFs are normalised by
          $1/\sigma$ where $\sigma$ is the standard deviation of the
          quantity plotted on the abscissa. Different curves
          correspond to different values of $r_{d}$ as shown in the
          insert. 
          The vertical line indicates the average 
          (a) $\langle (\widehat{\delta {\bm u}}\cdot \hat{\bm
            r})(\widehat{\delta {\bm f}_{p}}\cdot \hat{\bm r}) \rangle_{all}$ 
          and
          (b) $\langle (\widehat{\delta {\bm u}}\times \hat{\bm
            r})\cdot(\widehat{\delta {\bm f}_{p}}\times \hat{\bm r}) \rangle_{all}$. 
          $\langle Re_{\lambda} \rangle_{t}=178$ but these
          plots are very similar for $\langle Re_{\lambda}
          \rangle_{t}=80.9$.}
\label{fig:pdf_dot_cross}
\end{figure}

\begin{figure}
\centering
\subfigure{
        \begin{minipage}{0.7\linewidth}
                \includegraphics[clip,width=\linewidth]
                {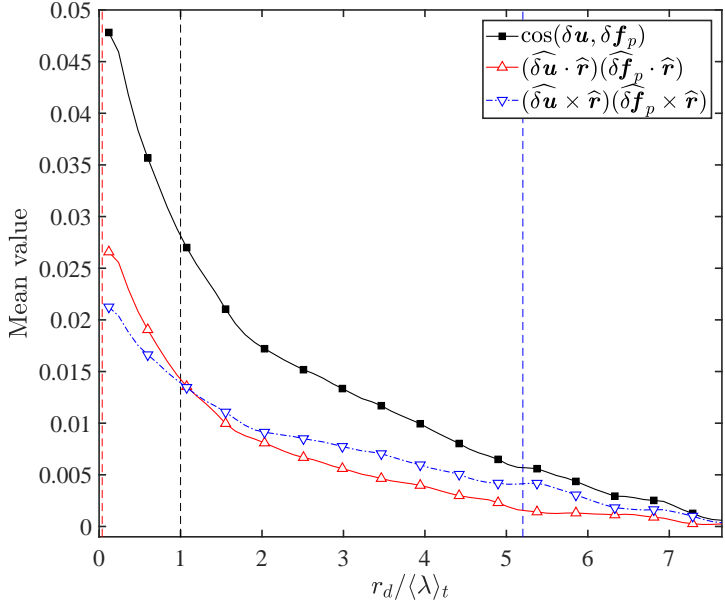}
        \end{minipage}
}
\caption{$\langle\cos\phi\rangle_{all}$, $\langle (\widehat{\delta
    {\bm u}}\cdot \hat{\bm r})(\widehat{\delta {\bm f}_{p}}\cdot
  \hat{\bm r})\rangle_{all}$ and $\langle (\widehat{\delta {\bm
      u}}\times \hat{\bm r})\cdot(\widehat{\delta {\bm f}_{p}}\times
  \hat{\bm r})\rangle_{all}$ as functions of $r_{d}/\langle \lambda
  \rangle_{t}$. The three dashed vertical lines indicate the positions
  on the abscissa of $\langle \eta \rangle_{t}$, $\langle \lambda
  \rangle_{t}$ and $\langle L\rangle_{t}$. $\langle Re_{\lambda}
  \rangle_{t}=178$ but this plot is very similar for $\langle
  Re_{\lambda} \rangle_{t}=80.9$.}
\label{fig:mean_dot_cross}
\end{figure}

Some further insight into the interactions between fluctuating
pressure gradient and fluctuating interscale energy transfer can be
gleaned by using
\begin{eqnarray}
\label{eq:cos_angle} 
\cos \phi = (\widehat{\delta {\bm u}}\cdot \hat{\bm
  r})(\widehat{\delta {\bm f}_{p}}\cdot \hat{\bm r}) +
(\widehat{\delta {\bm u}}\times \hat{\bm r})\cdot(\widehat{\delta {\bm
    f}_{p}}\times \hat{\bm r})
\:
\end{eqnarray}
where ${\bm f}_{p} \equiv -{\bm \nabla} p$, 
$\widehat{\delta {\bm f}_{p}}$ is the unit vector in the direction of 
$\delta {\bm f}_{p}$ and $\widehat{\delta {\bm u}}$ is the unit vector in the
direction of $\delta {\bm u}$.  The interest of the decomposition
(\ref{eq:cos_angle}) stems from the equality between the volume
integral $\iiint\limits_{|\bm{r}| \le r_d} 4\Pi(\bm{X},\bm{r},t) d{\bm
  r}$ and the surface integral $\iiint\limits_{|\bm{r}| = r_d} {\delta
  {\bm u}}\cdot \hat{\bm r} |{\delta {\bm u}}|^{2} d{\bm r}$ which
implies that compression events where $\delta {\bm u}\cdot \hat{\bm r}
<0$ contribute to the forward cascade from large to small scales
whereas stretching events where $\delta {\bm u}\cdot \hat{\bm r} >0$
contribute to backscatter, i.e. from small to large scales.  Hence,
the sign of $(\widehat{\delta {\bm u}}\cdot \hat{\bm
  r})(\widehat{\delta {\bm f}_{p}}\cdot \hat{\bm r})$ in the
decomposition (\ref{eq:cos_angle}) indicates whether the pressure
force does or does not act locally in the same way as the local
cascade event: $(\widehat{\delta {\bm u}}\cdot \hat{\bm
  r})(\widehat{\delta {\bm f}_{p}}\cdot \hat{\bm r})$ is positive when
the pressure force and the cascade are both compressing or stretching
and is negative when one of the two is compressing and the other
stretching.

The other term in the decomposition relates to energy exchanges from
one direction to another in ${\bm r}$-space and is therefore not
directly related to the cascade.  In figure~\ref{fig:pdf_dot_cross} we
plot the PDFs of the two terms in the right hand side of
(\ref{eq:cos_angle}). The overall shape of the PDF of $\cos\phi$ in
figure~\ref{fig:pdf_ang_du_dpg}(b) is reflected in the shape of the
PDF of $(\widehat{\delta {\bm u}}\times \hat{\bm
  r})\cdot(\widehat{\delta {\bm f}_{p}}\times \hat{\bm r})$ rather
than the PDF of $(\widehat{\delta {\bm u}}\cdot \hat{\bm
  r})(\widehat{\delta {\bm f}_{p}}\cdot \hat{\bm r})$.  Hence, the
wide variability in $\cos\phi$ values and also the gradual reduction
of this variability with decreasing length-scales $r_d$ within the
inertial and dissipation ranges seem to mostly reflect directional
rather than inter-scale effects in ${\bm r}$-space.
    
The mean values $\langle (\widehat{\delta {\bm u}}\cdot \hat{\bm
  r})(\widehat{\delta {\bm f}_{p}}\cdot \hat{\bm r})\rangle_{all}$ and
$\langle (\widehat{\delta {\bm u}}\times \hat{\bm
  r})\cdot(\widehat{\delta {\bm f}_{p}}\times \hat{\bm
  r})\rangle_{all}$ are both positive at all scales $r_d$ smaller than
$\langle L\rangle_{t}$, and are in fact increasing with decreasing
scale in this range (see figure~\ref{fig:mean_dot_cross}).  It is
clear that the small positive value of $\langle \cos\phi\rangle_{all}$
results from small positive values of both $\langle (\widehat{\delta
  {\bm u}}\cdot \hat{\bm r})(\widehat{\delta {\bm f}_{p}}\cdot
\hat{\bm r})\rangle_{all}$ and $\langle (\widehat{\delta {\bm
    u}}\times \hat{\bm r})\cdot(\widehat{\delta {\bm f}_{p}}\times
\hat{\bm r})\rangle_{all}$, the former being significantly larger than
the latter. The fact that $\langle (\widehat{\delta {\bm u}}\cdot
\hat{\bm r})(\widehat{\delta {\bm f}_{p}}\cdot \hat{\bm
  r})\rangle_{all}$ is small but positive means that there is a small
tendency for the pressure force to act in accordance with the local
interscale energy transfer on average, i.e. it has a compressing
effect when the interscale transfer is down the scales and a
stretching effect when the interscale transfer is up the scales. The
opposite happens too, but slightly less often. This small imbalance
creates an average picture for the terms in equation
(\ref{eq:cos_angle}) which, when combined with the negative average
value of $\Pi$, suggests an average interscale energy transfer from
large to small scales that is accompanied by a small compressive
effect by the pressure forces.

    The fluctuating picture, however, is once again significantly
    different as can be seen from the positive correlation between the
    fluctuations of $\cos\phi$ and the fluctuations of $\Pi$ in
    figure~\ref{fig:cor_coeff_ang_du_dpg}. We found similar positive
    correlations (as functions of $r_{d}/\langle \lambda\rangle_{t}$)
    between $(\widehat{\delta {\bm u}}\cdot \hat{\bm
      r})(\widehat{\delta {\bm f}_{p}}\cdot \hat{\bm r})$ and $\Pi$ on
    the one hand and between $(\widehat{\delta {\bm u}}\times \hat{\bm
      r})\cdot(\widehat{\delta {\bm f}_{p}}\times \hat{\bm r})$ and
    $\Pi$ on the other
    (not
    plotted here for economy of space). Even though the average of
    every term in equation (\ref{eq:cos_angle}) is positive (albeit
    small) and the average of $\Pi$ is negative (albeit also small
    compared to the fluctuations of $\Pi$), there is some correlation
    between negative/positive localised values of $\Pi$ and
    negative/positive localised values of $\cos\phi$,
    $(\widehat{\delta {\bm u}}\cdot \hat{\bm r})(\widehat{\delta {\bm
        f}_{p}}\cdot \hat{\bm r})$ and $(\widehat{\delta {\bm
        u}}\times \hat{\bm r})\cdot(\widehat{\delta {\bm f}_{p}}\times
    \hat{\bm r})$. The fluctuating picture is therefore, partly,
    opposite to the average one because it contains instances of
    events with negative $\Pi$ which correlate with negative values of
    every term in equation (\ref{eq:cos_angle}), in other words a
    significant number of cases where the pressure force has a
    stretching effect even though the interscale transfer is most
    likely down the scales. This is opposite to the average picture
    presented at the end of the previous paragraph.
    
\section{Conclusions}

The average picture of interscale energy transfers in statistically
stationary periodic turbulence is very different from the fluctuating
picture which hides behind it. 
Whereas only the interscale energy
transfer rate, the energy input rate, the turbulence dissipation rate
and the viscous diffusion in two-point separation space feature in the
average picture, the fluctuating picture contains intense and
intermittent fluctuations of all the terms in the KHMH equation
(\ref{eq:KHMH}), including the time-derivative, transport and
pressure-velocity terms, all of which average out to zero. In fact the
most intense fluctuations are those of the time-derivative and the
transport term, followed closely by the fluctuations of the interscale
energy transfer rate and the pressure-velocity term 
(figure~\ref{fig:deviations}). 
The fluctuations of each one of these four terms are 
much more intense than the fluctuations of the turbulence dissipation 
rate which is known to be highly intermittent since the 1950s 
\citep{batchelor_1949_the, kolmogorov_1962_a,
  vincent_1991_the,frisch_1995_turbulence}.

More importantly, perhaps, many of these fluctuations are
correlated. These correlations result from the sweeping effect and the
link between non-linearity and non-locality. The sweeping effect
introduces correlations between the time-derivative term
$\mathcal{A}_t$ and terms having their origin in the convective
non-linearity of the Navier-Stokes equation, namely the transport term
$\mathcal{T}$, the interscale energy transfer rate (with a minus sign
for a positive correlation) $-\mathcal{\Pi}$ and
$\mathcal{T}-\mathcal{\Pi}$. It is important to realise that the
interscale energy transfer and related cascade dynamics expressed by
$\Pi$ decorrelate from the sweeping effect with decreasing
length-scales and increasing Reynolds number. In these limits, the
sweeping effect's partial cancellation is assumed mainly or
exclusively by the correlation between 
$\mathcal{A}_t$ and $\mathcal{T}$. 

The link between non-linearity and non-locality which exists in
incompressible Navier-Stokes flow is expressed in the significant
correlations between the pressure-velocity term $\mathcal{T}_p$ and
each of $\mathcal{\Pi}$, $\mathcal{T}$ and
$\mathcal{\Pi}-\mathcal{T}$. The sweeping and the non-linear/non-local
link effects are separate in that $\mathcal{A}_t$ and $\mathcal{T}_p$
are uncorrelated at all scales. In the limit of small length scales
larger than the average Taylor micro-scale (i.e. within what might be
termed the inertial range at high enough Reynolds number values) and
also as the Reynolds number grows, the correlation with the pressure
fluctuations via $\mathcal{T}_p$ is increasingly with the interscale
transfer rate $\mathcal{\Pi}$ and diminishingly with the transport
term $\mathcal{T}$. The fluctuating pressure has therefore an
important role to play in the interscale energy transfer and the
energy cascade.

The role of the two-point fluctuating pressure force difference on
energy exchanges in ${\bm r}$-space and interscale energy transfers in
particular is manifest in the geometrical alignments that it has with
the two-point fluctuating velocity difference. These alignments can be
decomposed in two parts (see equation \ref{eq:cos_angle}), one
directly concerned with interscale, including cascade, events and the
other with reorientation events in ${\bm r}$-space. The fluctuating
pressure force is involved in both processes and can have a
compressing or a stretching effect on either backscatter or forward
cascade events in a way that contributes to correlation statistics and
even mean values such as those of figure~\ref{fig:mean_dot_cross}.

It is clear that the average picture of the cascade and interscale
exchanges in ${\bm r}$-space does not represent much of the actual
physical processes involved even in statistically stationary periodic
turbulence. It makes little sense to try to model an unrepresentative
average picture rather than attempt to directly model the underlying
physics which appear when the clouding effects of averages are
lifted. In the case of Large Eddy Simulations, for example, it will be
important to supplement existing subgrid modelling approaches, which
may have some realism in only part of the flow, with models based on
conditional statistics and cross-scale correlations involving various
processes, including fluctuating pressure forces. Our results
concerning fluctuations, intermittency and the presence of relatively
rare but powerful events also suggest the need for further studies of
two-point correlations between fluctuating velocity gradients,
pressure gradients and fluctuating velocities. Such studies directly
impact on correlations between fluctuating velocities and fluctuating
pressure forces which need to be modelled in Reynolds-Stress-Transport
models.

\section*{Acknowledgements}

We thank Professor Susumu Goto for letting us use his parallelised
pseudo-spectral DNS code for periodic turbulence. 
We also acknowledge support from ERC Advanced Grant 320560. 
%

\bibliographystyle{jfm}

\end{document}